\documentclass[cup6b]{caps}
\usepackage{graphics}
\newcommand{\la}{\;\raisebox{-0.3ex}{\mbox{$\stackrel{<}{_\sim} \;$}}}
\newcommand{\ga}{\;\raisebox{-0.3ex}{\mbox{$\stackrel{>}{_\sim} \;$}}}

\begin{document}

\pagenumbering{roman}
\setcounter{chapter}{15}

\pagenumbering{arabic}

\author[T. M. Tauris$\;$ and $\,$ E. P. J. van den Heuvel]{T. M. Tauris\\
        Astronomical Observatory, Niels Bohr Institute, Copenhagen University, Denmark
\and    E. P. J. van den Heuvel\\
        Astronomical Institute, University of Amsterdam, The Netherlands}
\chapter[Formation and Evolution of Compact Stellar {X}-ray Sources]
        {Formation and Evolution of Compact\\Stellar {X}-ray Sources}

\section{Introduction and brief historical review}
In this chapter we present an overview of the formation and evolution of compact stellar
{X}-ray sources. For earlier reviews on the subject we refer to Bhattacharya \& van~den~Heuvel~(1991),
van~den~Heuvel~(1994) and Verbunt \& van~den~Heuvel~(1995). 

In our galaxy there are about 100 bright {X}-ray sources with
fluxes well above $10^{-10}$~erg$\,$cm$^{-2}\,$s$^{-1}$ in the energy range
$1-10$~keV (above the Earth's atmosphere). The distribution of these
sources shows a clear concentration towards the Galactic center and also
towards the Galactic plane, indicating that the majority do indeed belong to
our galaxy. Furthermore, a dozen strong sources are found in Galactic
globular clusters and in the Magallanic Clouds. Shortly after the discovery of the first source
(Sco~X-1, Giacconi~et~al.~1962) Zel'Dovitch \& Guseinov~(1966), Novikov \& Zel'Dovitch~(1966) and
Shklovskii (1967) suggested that the strong Galactic {X}-ray sources are
accreting neutron stars or black holes in binary systems. (The process of mass accretion onto a
supermassive black hole had already a few years earlier been suggested as the energy source for
quasars and active galactic nuclei by Salpeter~(1964), Zel'Dovitch~(1964) and Zel'Dovitch \& Novikov~(1964).)

The {X}-ray fluxes measured correspond to typical
source luminosities of $10^{34}-10^{38}$~erg$\,$s$^{-1}$
(which is more than 25$\,$000 times the total energy output of our Sun).
Table~\ref{stellar_obj_acc} lists the rates of accretion required to generate a
typical {X}-ray luminosity of $10^{37}$~erg$\,$s$^{-1}$. Also listed
is the amount of gravitational potential energy released per unit
mass ($\Delta U/m = GM/R$) by accretion onto a $1\,M_{\odot}$ stellar (compact) object,
as well as the column density towards the stellar surface
(or Schwarzschild radius) in the case of spherical accretion,
$\sigma = L_x\,4\pi\,\sqrt{R/(GM)^3}$.
The table shows that only for accreting neutron stars and black holes
is the column density low enough to allow {X}-rays to escape,
as {X}-rays are stopped at column densities larger than a few
g$\,$cm$^{-2}$. Hence, the strongest Galactic sources cannot be
accreting white dwarfs.
 \begin{table*}
  \caption{Energetics of accretion -- see text $\qquad(\;^*L_x=10^{37}$~erg/s)}
    \begin{tabular}{l|rlccc}
     \hline \hline
     Stellar object & Radius & $\Delta U/mc^2$ & $\Delta U/m$ & $dM/dt\;^*$ & Column density$\;^*$\\
     $1\,M_{\odot}$ & (km) & & (erg/g) & ($M_{\odot}$/yr) & (g/cm$^2$)\\
     \hline
     Sun & $7\times 10^{5}$ & $2\times 10^{-6}$ & $2\times 10^{15}$ & $1\times 10^{-4}$ & 140\\
     White dwarf & $10\,000$ & $2\times 10^{-4}$ & $1\times 10^{17}$ & $1\times 10^{-6}$ & 16\\
     Neutron star& 10 & 0.15 & $1\times 10^{20}$ & $1\times 10^{-9}$ & 0.5\\
     Black hole & 3 & $0.1 \sim 0.4$ & $4\times 10^{20}$ & $4\times 10^{-10}$ & 0.3\\
     \hline \hline
    \end{tabular}
  \label{stellar_obj_acc}
 \end{table*}

The accreting neutron star binary model was nicely confirmed
by Schreier et~al. (1972) who discovered that the source
Cen~X-3 is regularly pulsing (thus: a neutron star),
and is member of an eclipsing binary system. Its regular {X}-ray
pulsations have a period of 4.84~sec, and the regular {X}-ray eclipses
have a duration of 0.488~days, which repeat every 2.087~days.
The pulse period shows a sinusoidal doppler modulation with the
same 2.087~day period and is in phase with the {X}-ray eclipses,
indicating that the {X}-ray pulsar is moving in a circular orbit
with a projected velocity of 415.1~km~s$^{-1}$.
Hence, the 4.84~sec period of the {X}-ray pulsations is the
rotation period of the neutron star and the binary orbital period
is 2.087~days.
Using Kepler's~3.~law ($\Omega ^2 = GM/a^3$) combined with the eccentricity and measured
radial velocity amplitude ($K_x=\Omega\,a_x\sin i/\sqrt{1-e^2}$) of the {X}-ray pulsar
one can determine the so-called mass function of the binary:
\begin{equation}
  f(M) = \frac{M_2^3\,sin^3\,i}{(M_X + M_2)^2} = \frac{1}{2\pi G}\;K_x^3\,P_{\rm orb}\,(1-e^2)^{3/2}
\label{massfunction}
\end{equation}
where $M_X$ and $M_2$ denote the masses of the accreting compact star
and its companion star, respectively, and $i$ is the inclination of the orbital
angular momentum vector with respect to the line-of-sight to the Earth. For Cen~X-3 the mass function is
$f=15.5\,M_{\odot}$ and thus one can derive
a minimum companion mass of about $18\,M_{\odot}$.\\
In Cyg~X-1 a bright 09.7 supergiant star was identified as the optical counterpart
(Webster \& Murdin 1972; Bolton 1972). Based on its large velocity amplitude (72~km$\,$s$^{-1}$)
this {X}-ray source, with a  5.6~day orbital period, must host a black hole given the fact that
the derived mass of the compact object $> 3\,M_{\odot}$ when assuming a realistic mass
($\ge 15\,M_{\odot}$) for the O9.7 supergiant.

The recognition that neutron stars and black holes can exist in close binary systems came at first
as a surprise. It was known that the initially more massive star should evolve first and explode
in a supernova (SN). However, as a simple consequence of the virial theorem, the orbit of the
post-SN system should be disrupted if more than half of the total mass of the binary is suddenly ejected
(Blaauw~1961). For {X}-ray binaries like Cen~X-3 it was soon realized (van den Heuvel \& Heise 1972;
and independently also by Tutukov \& Yungelson 1973)
that the survival of this system was due to the effects of large-scale mass transfer that must have occurred
prior to the SN. 

The formation of low-mass {X}-ray binaries ($M_{\rm donor} \le 1.5\,M_{\odot}$)
with observed orbital periods mostly between 11~min. and
12~hr., as well as the discovery of the double neutron star system PSR~1913+16 (Hulse \& Taylor 1975)
with an orbital period of 7.75~hr., was a much tougher nut to crack. How could these stars end up being
so close when the progenitor star of the neutron star must have had a radius much larger than the current separation$\,$?
It was clear that such systems must have lost a large amount of orbital angular momentum. The first models
to include large loss of angular momentum were made by van~den~Heuvel \& de~Loore~(1973) for the later
evolution of HMXBs, showing that in this way very close systems like Cyg~X-3 can be formed, and Sutantyo~(1975)
for the origin of LMXBs.
The important concept of a `common envelope' (CE) evolution was introduced by Paczy\'nski~(1976) and Ostriker~(1976).
This scenario, as well as that of van~den~Heuvel \& de~Loore~(1973), 
could link the massive {X}-ray binary Cyg~X-3 and the binary radio pulsar PSR~1913+16.
In this scenario a neutron star is captured by the expansion of a giant companion star and 
is forced to move through the giant's envelope.
The resulting frictional drag will cause its orbit to shrink rapidly while, at the same time,
ejecting the envelope before the naked core of the giant star explodes to form another neutron star.
It was suggested by Smarr \& Blandford~(1976) that it is an old `spun-up' neutron star which
is observed as a radio pulsar in PSR~1913+16. The magnetic field of this visible pulsar is relatively
weak ($\sim\!10^{10}$ Gauss, some two orders of magnitude lower than the average pulsar
magnetic field) and its spin period is very short (59~ms). Hence, this pulsar is most
likely spun-up (or `recycled') in an {X}-ray binary where mass and angular momentum from an
accretion disk is fed to the neutron star (as already suggested by Bisnovatyi-Kogan \& Komberg 1974). 
The other neutron star in the system was then
produced by the second supernova explosion and must be a young, strong $\vec{B}$--field neutron star
(it is not observable -- either because it has already rapidly spun-down, due to
dipole radiation, or because the Earth is not hit by the pulsar beam).\\ 
The idea of recycling pulsars was given a boost by the discovery of the first 
millisecond radio pulsar in 1982 (Backer~et~al.~1982). As a result of the long accretion phase
in low-mass {X}-ray binaries, millisecond pulsars are believed to be formed in such systems
(Alpar~et~al.~1982; Radhakrishnan \& Srinivasan~1982). This model was beautifully confirmed
by the discovery of the first millisecond {X}-ray pulsar in the low-mass {X}-ray binary system
SAX~1808.4--3658 (Wijnards \& van~der~Klis~1998). 
Now already four of these accreting millisecond pulsars are known (Galloway et~al. 2002; Markwardt, Smith \& Swank 2003).
For a detailed discussion on the
evidence for the presence of rapidly spinning weakly magnetized neutron stars in low-mass
{X}-ray binaries we refer to the chapters on `{\em QPO observations and theories}' by van~der~Klis 
and `{\em {X}-ray bursts}' by Strohmayer \& Bildsten presented elsewhere in this book.\\
Finally, another ingredient which has important consequences for close binary evolution is the event of a
`kick' imparted to newborn neutron stars as a result of an asymmetric SN. They were first applied to
binary systems by Flannery \& van~den~Heuvel (1975). There is now ample evidence for the occurrence of
such kicks inferred from the space velocities of pulsars and from dynamical effects on surviving binaries
(e.g. Dewey \& Cordes~1987; Lyne \& Lorimer~1994 and Kaspi~et~al.~1996). Furthermore, as argued by Kalogera \& Webbink~(1998),
without kicks the formation of LMXBs with short orbital periods cannot be explained.

In this review we will concentrate on the stellar evolutionary processes which are responsible
for the formation of the different types of compact binaries and their evolution.
Our focus is mainly on binaries with neutron star or black hole accretors. We refer to other reviews 
for discussions on CVs and AM~CVn systems (Warner \& Kuulkers, this book), super-soft sources (Kahabka, this book)
and sdB-star binaries (Maxted~et~al 2001; Han~et~al.~2002).
In the next section, however, we will first give a short introduction to the observational properties of the
{X}-ray binaries and binary pulsars which we discuss afterwards. A more general introduction to {X}-ray binaries is
given by Chakrabarty \& Psaltis in the beginning of this book.

\section{Compact binaries and their observational properties}
\begin{figure}
  \centering
    \centerline{\resizebox{8cm}{!}{\includegraphics{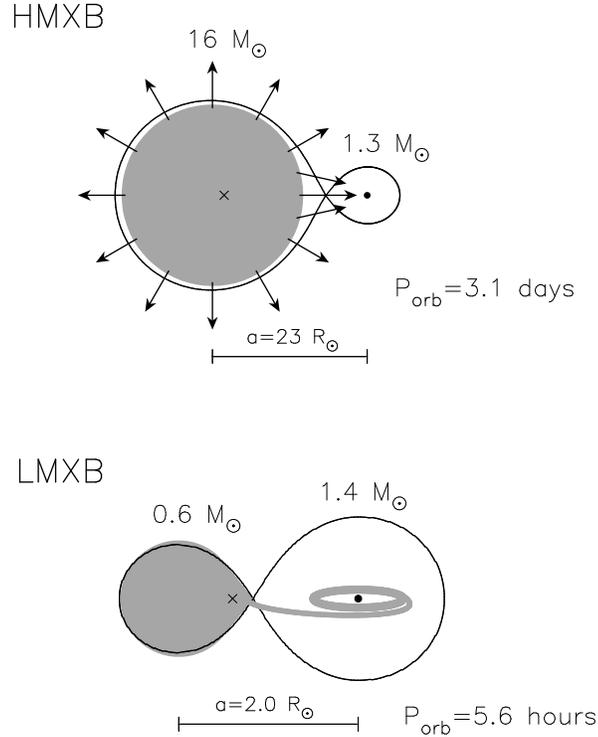}}}
  \caption{Examples of a typical HMXB (top) and LMXB (bottom). The neutron star in the HMXB is fed by a
           strong high-velocity stellar wind and/or by beginning atmospheric Roche-lobe overflow.
           The neutron star in an LMXB is surrounded by an accretion disk which is fed by Roche-lobe overflow.
           There is also observational evidence for HMXBs and LMXBs harbouring black holes.}
  \label{Xbinary}
\end{figure}

Over 90\% of the strong Galactic {X}-ray sources appear to fall into two distinct groups:
the high-mass {X}-ray binaries (HMXBs) and the low-mass {X}-ray binaries (LMXBs).
These two groups differ in a number of physical characteristics 
(see Fig.~\ref{Xbinary} and Table~\ref{2classes} for some examples).
As we shall see later on, binary pulsars and single millisecond pulsars are the descendants of the 
{X}-ray binaries containing an accreting neutron star. 

\subsection{High-mass {X}-ray binaries -- HMXBs}
\begin{figure}[h]
  \centering
    \centerline{\resizebox{9cm}{!}{\includegraphics{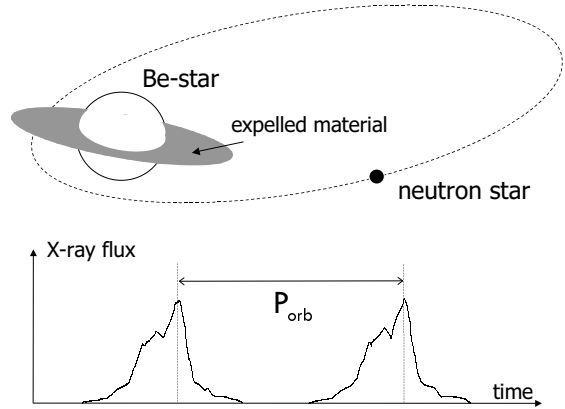}}}
  \caption{Schematic model of a Be-star X-ray binary system. The neutron star moves in an eccentric orbit
           around the Be-star which is not filling its Roche-lobe. However, near the periastron passage
           the neutron star accretes circumstellar matter, ejected from the rotating Be-star, resulting in
           an X-ray burst lasting several days.}
  \label{Be-star}
\end{figure}

 \begin{table*}[t]
  \caption{The two main classes of strong Galactic {X}-ray sources}
    \begin{tabular}{lll}
     \hline \hline
     & HMXB & LMXB\\
     \hline
     {X}-ray spectra: & $kT \ge$ 15 keV (hard) & $kT \le$ 10 keV (soft)\\
     \noalign{\smallskip}
     Type of time variability: & regular {X}-ray pulsations & only a very few pulsars\\ 
                               & no {X}-ray bursts & often {X}-ray bursts\\
     \noalign{\smallskip}
     Accretion process: & wind (or atmos. RLO) & Roche-lobe overflow\\
     \noalign{\smallskip}
     Timescale of accretion: & $10^5$ yr & $10^7 - 10^9$ yr\\
     \noalign{\smallskip}
     Accreting compact star: &  high $\vec{B}$-field NS (or BH) & low $\vec{B}$-field NS (or BH) \\
     \noalign{\smallskip}
     Spatial distribution: & Galactic plane & Galactic center and\\
                           &                & spread around the plane\\
     \noalign{\smallskip}
     Stellar population:  & young,  age $< 10^7$ yr & old, age $> 10^9$ yr\\
     \noalign{\smallskip}
     Companion stars: & luminous, $L_{\rm opt}/L_x >1$ & faint, $L_{\rm opt}/L_x \ll 0.1$\\
                      & early-type O(B)-stars & blue optical counterparts\\
                      & $> 10\,M_{\odot}$ (Pop.$\,$I) & $\le 1\,M_{\odot}$ (Pop.$\,$I and II)\\
     \hline \hline
    \end{tabular}
  \label{2classes}
 \end{table*}
There are about 130 known HMXBs (Liu, van~Paradijs \& van~den~Heuvel 2000) and 25 have well-measured
orbital parameters.
There are $\sim\!40$ pulsating HMXB sources with typical pulse periods between $10-300$~seconds
(the entire observed range spans between 0.069~seconds and 20~minutes).
Among the systems with $P_{\rm orb} \le 10$~days and $e\le 0.1$ are the strong sources and
`standard' systems such as Cen~{X}-3 and SMC~{X}-1. These are
characterized by the occurrence of regular {X}-ray eclipses and double-wave ellipsoidal light
variations produced by tidally deformed (`pear-shaped') giant or sub-giant companion stars with masses $> 10\,M_{\odot}$. 
However, the optical luminosities ($L_{\rm opt} > 10^5\,L_{\odot}$) and spectral types
of the companions indicate original ZAMS masses $\ge 20\,M_{\odot}$, corresponding to O-type progenitors.
The companions have radii $10-30\,R_{\odot}$ and (almost) fill their critical Roche-lobes, see Sect.~\ref{RLO}. 
In a number of pulsating sources, such as X0115+63 (and Her~X-1, an intermediate-mass {X}-ray binary system) there are absorption/emission 
features in the {X}-ray spectrum which
are most probably cyclotron lines, resulting from magnetic fields with strengths $B\simeq 5\times 10^{12}$~G
(Kirk \& Tr\"{u}mper~1983).
Among the standard HMXBs, there are at least two
systems that are thought to harbour black holes: Cyg~X-1 and LMC~X-3.\\
Another group of HMXBs consists of the moderately wide, eccentric binaries with $P_{\rm orb}\simeq 20-100$~days
and $e\simeq 0.3-0.5$. A new third (sub-)group has recently been proposed by Pfahl~et~al.~(2002). These systems
have $P_{\rm orb}\simeq 30-250$~days and small eccentricities $e\la 0.2$. Together these two groups
form a separate sub-class of HMXBs: the Be-star {X}-ray binaries 
(see Fig.~\ref{Be-star}; first recognized as a class by Maraschi, Treves \& van~den~Heuvel 1976).
In the Be-star {X}-ray binaries the companions are
rapidly rotating B-emission stars situated on, or close to, the main-sequence (luminosity class III--V).
There are more than 50 such systems known making them the most numerous class of HMXBs
(see, for example, van~den~Heuvel \& Rappaport 1987 and Chakrabarty \& Psaltis in this book for a review).
The Be-stars are deep inside their
Roche-lobes, as is indicated by their generally long orbital periods ($\ga 15$ days) and by the absence
of {X}-ray eclipses and of ellipsoidal light variations. According to the luminosities and spectral
types, the companion stars have masses in the range about $8-20\,M_{\odot}$ (spectral types O9--B3, III--V).
The {X}-ray emission from the Be-star {X}-ray systems tends to be extremely variable, ranging from complete
absence to giant transient outbursts lasting weeks to months. 
During such an outburst episode one often observes orbital modulation of the {X}-ray emission, due to
the motion of the neutron star in an eccentric orbit, see Fig.~\ref{Be-star}.
The recurrent {X}-ray outbursts are most probably related to the
irregular optical outbursts generally observed in Be-stars, which indicate sudden outbursts of
mass ejection, presumably generated by rotation-driven instability in the equatorial regions of
these stars (see, for example,  Slettebak 1988). While the Be-star {X}-ray binaries are transient sources
(often unobservable for months to years) the `standard' systems are persistent {X}-ray sources.
HMXBs are located along the Galactic plane among their OB-type progenitor stars.

\subsection{Low-mass {X}-ray binaries -- LMXBs}
Orbital periods have been measured for some 30 of these systems. They range from 11~min. to 17~days,
similar to the orbital periods of cataclysmic variables -- see review by Charles elsewhere in this book. Only in the few
widest of these systems one can observe the spectrum of the optical companion. In all other systems,
the optical spectrum is that of the hot accretion disk.
The LMXBs are very seldom {X}-ray pulsars. The reason is their relatively weak magnetic fields
$10^9\sim 10^{11}$~G, which is expected to result from accretion-induced field decay (Taam \& van~den~Heuvel 1986; Geppert \& Urpin 1994; 
Konar \& Bhattacharya 1997; Bhattacharya 2002; Cumming, Zweibel \& Bildsten 2001). On the other hand these sources show {X}-ray bursts
(sudden thermonuclear fusion of accreted matter at the surface of the neutron star -- see review by
Strohmayer \& Bildsten in this book), which are suppressed if the magnetic field strength is $>10^{11}$~G (Lewin \& Joss 1983).
For this reason such bursts are not observed in HMXBs.
The discovery of (kilo-Hertz) quasi-periodic oscillations (QPOs) in the {X}-ray flux of LMXBs
has provided a clear timing signature of the accreting neutron stars and black holes
in these systems. In the past decade much insight of detailed accretion physics and
testing of general theory of relativity
has been revealed by observations and 'beat frequency' models (see van~der~Klis for a review
elsewhere in this book).
There is more than a dozen of LMXBs systems for which there is strong
evidence for the presence of a black hole -- see Sect.~\ref{SXT} and the review by Charles in this book. 

Most of the LMXBs are located in the Galactic bulge and in globular clusters, and thus appear to 
belong to an old stellar population. They do not show considerable run-away characteristics, as they
are confined mostly to within 0.6~kpc of the Galactic plane -- however some LMXBs have transverse
velocities in excess of 100~km$\,$s$^{-1}$. It is interesting to notice that the systems in globular clusters 
must have velocities
smaller than the escape velocities from these clusters, which less than about 30~km$\,$s$^{-1}$.

\subsection{Intermediate-mass {X}-ray binaries -- IMXBs}
Above we have described the HMXB and LMXB systems with companion stars $> 10\,M_{\odot}$ and $\la 1\,M_{\odot}$, respectively.
There must exist also a large number of Galactic compact binaries with
companion star masses in the interval $1-10\,M_{\odot}$ as has been argued e.g. by van~den~Heuvel~(1975). 
These are the so-called intermediate-mass {X}-ray binaries 
(IMXBs). IMXBs have recently been recognized as a class of their own. 
It is important to be aware that IMXB systems are not easily observed as a result of
a simple selection effect against {X}-ray sources with intermediate-mass companions.
The reason is the following (van~den~Heuvel 1975): `Standard' HMXBs have evolved (sub)giant companions which are massive enough to
have a strong stellar wind mass-loss rate (typically $\dot{M}_{\rm wind} \simeq 10^{-6}\, M_{\odot}$~yr$^{-1}$) sufficient to power
a bright {X}-ray source, via an accreting neutron star or black hole, for $10^5 - 10^6$~yr. 
The LMXBs are not wind-fed {X}-ray sources. These systems experience mass transfer via Roche-lobe
overflow (RLO) from their companion star to the compact object. The LMXBs often evolve slowly on a
nuclear timescale ($10^8 \sim 10^9$ yr) and the majority of the transferred material is usually funneled onto
the compact object, via an accretion disk, yielding accretion rates of $10^{-10}-10^{-8}\,M_{\odot}$~yr$^{-1}$.
In IMXBs the companions are not massive
enough to produce sufficiently high wind mass-loss rates to power an observable {X}-ray source.
Subsequently when IMXBs evolve through RLO the relatively large mass ratio between the companion star 
and a neutron star causes this phase to be short lived -- either on a subthermal timescale (Tauris, van~den~Heuvel \& Savonije~2000),
or the system evolves through a common envelope. In either case, the systems
evolve on a timescale of only a few 1000~yr. Furthermore, the very high mass-transfer rates under these circumstances
($\dot{M} > 10^{-4}\,M_{\odot}$~yr$^{-1} \gg \dot{M}_{\rm Edd}$) may cause the emitted {X}-rays to be absorbed 
in the dense gas surrounding the accreting neutron star.
We therefore conclude that HMXBs and LMXBs are naturally selected as
persistent {X}-ray sources fed by a strong stellar wind and RLO, respectively.

Also, from a theoretical point of view we know that IMXBs must exist in order to
explain the formation of binary pulsars with heavy {CO} or {ONeMg} white dwarf companions
(Tauris, van~den~Heuvel \& Savonije~2000). Despite the above mentioned selection effects against IMXBs there
are a few such systems with neutron stars detected:
Her~X-1 and Cyg~X-2 are systems of this type. In the latter system the companion presently has a
mass $< 1\,M_{\odot}$, but it is highly overluminous for this mass, which indicates that it is an evolved
star that started out with a mass between 3 and $4\,M_{\odot}$ at the onset of the mass-transfer phase
(Podsiadlowski \& Rappaport 2000; King \& Ritter 1999).
Among the black hole {X}-ray binaries, IMXBs are more common, for example: 
GRO~J1655--40 ($M_{\rm d}\sim\!1.5\,M_{\odot}$; Beer \& Podsiadlowski 2002), 4U$\,$1543--47 ($M_{\rm d}\sim\!2.5\,M_{\odot}$;
Orosz et~al. 1998), LMC~X-3 ($M_{\rm d}\sim\!5\,M_{\odot}$; Soria et~al. 2001) 
and V$\,$4642~Sgr ($M_{\rm d}\sim\!6.5\,M_{\odot}$; Orosz et~al. 2001). In these systems
the donor star ($M_{\rm d}$) is less massive than the black hole, so mass transfer by Roche-lobe overflow is stable.

\subsection{Soft {X}-ray transients -- SXTs}
\label{SXT}
A great breakthrough in the discovery of black holes in {X}-ray binaries 
came with the discovery by McClintock \& Remillard~(1986) that the
K5V companion of the source A0620--00, in a spectroscopic binary with $P_{\rm orb}=7.75\,$ hr,
has a velocity amplitude $> 470$ km$\,$s$^{-1}$. This large orbital velocity indicates that
even if the K-dwarf would have zero mass, the compact object has a mass $> 3\,M_{\odot}$
and therefore must be a black hole. Since then a dozen such systems consisting of a black hole
with a low-mass donor star have been discovered.
These black hole systems are the so-called soft {X}-ray transients (SXT)
and appear as bright {X}-ray novae with luminosities $L_x \sim\!10^{38}$~erg$\,$s$^{-1}$ for several weeks.
At the same time, the optical luminosity of these systems brightens by 6 to 10 magnitudes, making
them optical novae as well. After the decay of the {X}-ray and optical emission, the spectrum of a
K or G~star becomes visible, and the large amplitude variations ($\ge 100$~km$\,$s$^{-1}$) in the radial
velocities of these stars indicate a compact object mass of $>3\,M_{\odot}$ (exceeding the
maximum mass thought possible for neutron stars). Their orbital 
periods are between 8 hr. and 6.5 days (see e.g. Lee, Brown \& Wijers 2002 and McClintock \& Remillard in this book). 

\subsection{Peculiar {X}-ray binaries}
\begin{figure}
  \centering
 \centerline{\resizebox{23.0cm}{!}{\includegraphics{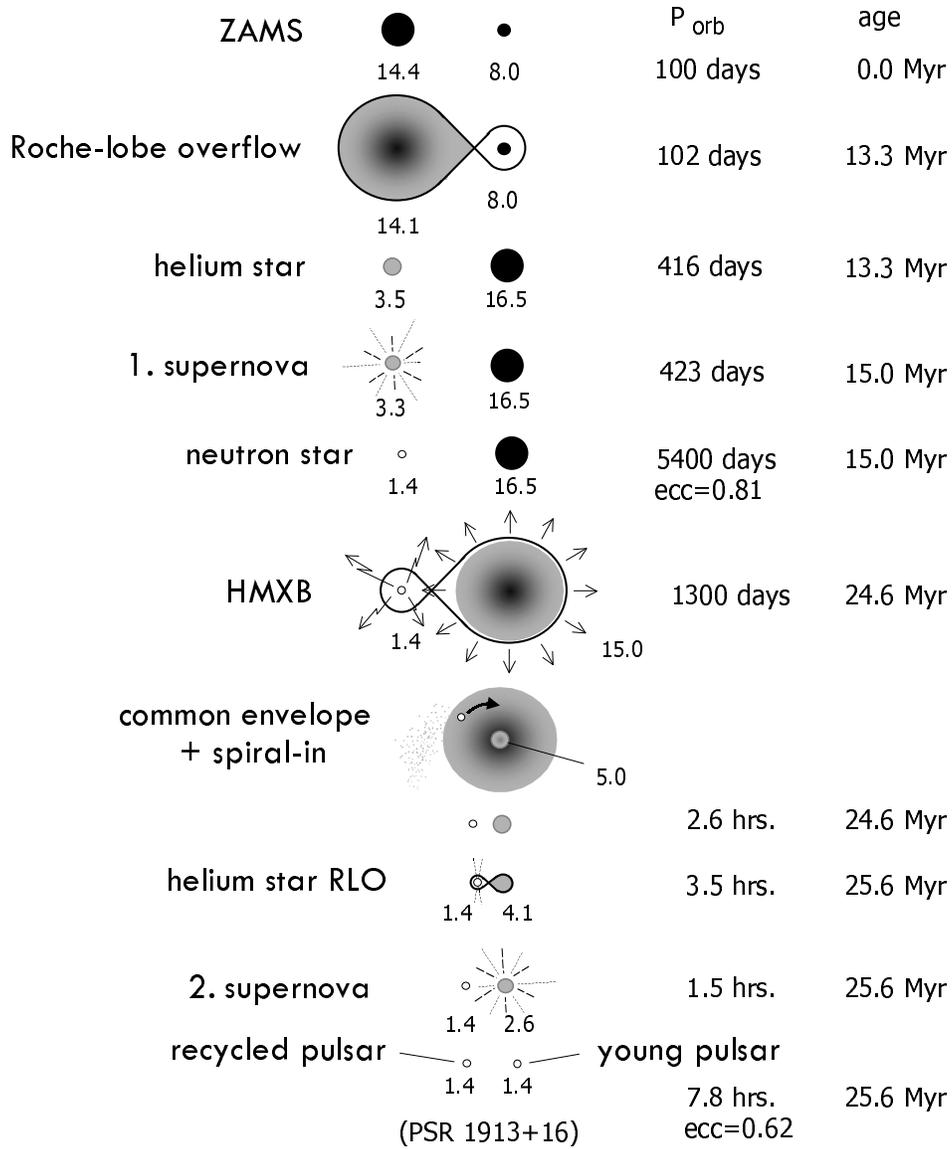}}}
  \caption{Cartoon depicting the formation of a Be-star/HMXB and finally 
           a double neutron star system. Such a binary will experience two supernova
           explosions. It is always the recycled pulsar which is observed in a double pulsar system as a result of
           its very long spin-down timescale compared to the young pulsar (a factor of $\sim\!10^2$).
           Tight NS-NS systems will coalesce due to gravitational wave radiation. These collisions should be
           detected by advanced gravitational wave detectors such as LIGO~II/VIRGO.
           }
  \label{HMXB-cartoon}
\end{figure}

Not all observed {X}-ray binaries fall into the well defined classes described above.
Among the more intriguing systems are SS433 and Cyg~X-3 which both have flaring radio emissions and jets
(see e.g. review by Fender in this book). Cyg~X-3 ($P_{\rm orb}=4.8$~hr) is probably a later evolutionary
phase of a wide HMXB (e.g. see Fig.~\ref{HMXB-cartoon} in Sect.~\ref{HMXB_evol}; van~den~Heuvel \& de~Loore 1973). 

\subsection{The binary and millisecond radio pulsars}
\begin{figure}
  \centering
    \centerline{\resizebox{12cm}{16.8cm}{\includegraphics{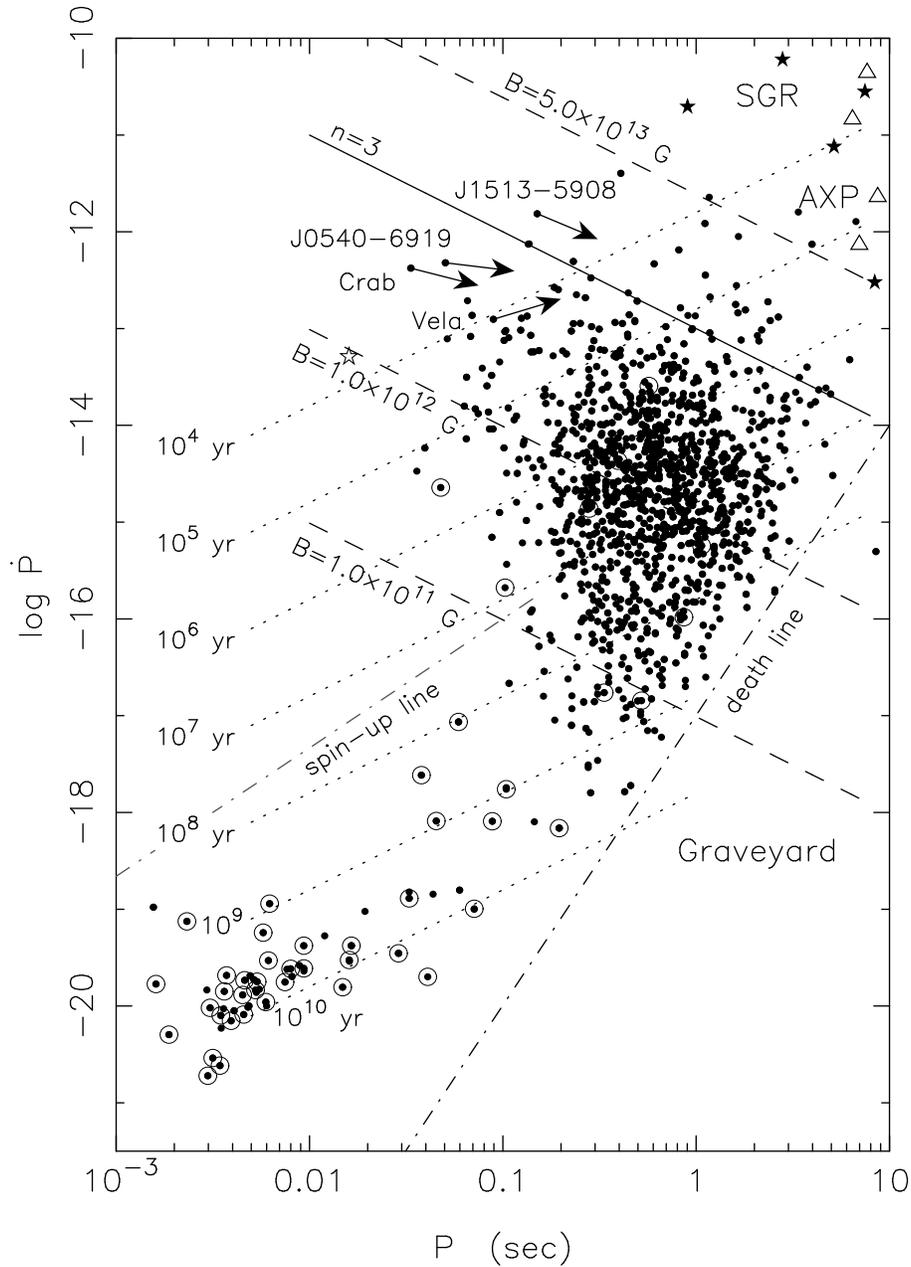}}}
  \caption{($P,\dot{P}$)-diagram of $\sim\! 1300$ observed radio pulsars (ATNF Pulsar Catalogue data).
            Binary pulsars are marked by a circle.
            Soft gamma-ray repeaters (SGR) and anomalous X-ray (AXP) pulsars are marked by 
            stars and triangles, respectively. Also shown are lines of constant surface dipole magnetic
            field strength (dashed) and characteristic ages (dotted). The arrows marked on a few young
            pulsars indicate a measurement of the braking index. The ``death line'' is the pair-creation limit
            for generating radio pulses.
           }
  \label{PPdot}
\end{figure}
The about 50 Galactic binary radio pulsars detected so far are generally characterized by short spin periods, $P$ and
small values of the period derivative, $\dot{P}$. This can clearly be seen in a ($P,\dot{P}$)-diagram
of radio pulsars, see Fig.~\ref{PPdot}. Simple expressions for the surface magnetic dipole field strength, $B$ and
`spin-down' age, $\tau$ are given by the magnetic dipole
model of pulsars (see Manchester \& Taylor 1977 for further details):
\begin{equation}
  B = \sqrt{\frac{3c^3I }{8\pi ^2 R^6}\,P\dot{P}} \simeq 3 \times 10^{19}\,\sqrt{P\dot{P}}\quad{\rm Gauss}
\end{equation}
where $I$ and $R$ are the moment of inertia ($\sim\!10^{45}$~g$\,$cm$^2$) and radius of the neutron star, respectively, and
\begin{equation}
  \tau \equiv P/2\dot{P}
\end{equation}
Fig.~\ref{PPdot} shows that the binary pulsars typically have $B= 10^8 - 10^{10}$~G and
$\tau =$ 1$\,$--$\,$10~Gyr, whereas the ordinary isolated pulsars have  $B= 10^{12} - 10^{13}$~G
and $\tau \la 10\,$Myr. As will be discussed in a later section in this chapter, the short spin periods 
and low values of $B$ are consequences of the recycling process where an old neutron star
accretes mass and angular momentum from its companion star via an accretion disk. 
For theoretical calculations of the dynamical evolution of single pulsars in the ($P,\dot{P}$)-diagram 
we refer to Tauris \& Konar~(2001).\\
There are four classes of binary pulsars detected so far (see Table~\ref{binaries}). 
The different classes are: i) high-mass binary pulsars (HMBPs) with a neutron star or
{ONeMg}/{CO} white dwarf companion, ii) low-mass binary pulsars (LMBPs) with
a helium white dwarf companion, iii) non-recycled pulsars with a {CO} white dwarf companion,
and finally, iv) pulsars with an unevolved companion. A few further sub-divisions can be introduced
from an evolutionary point of view.
The globular cluster pulsars will not be discussed here since these systems are the result of
tidal capture or exchange encounters -- see Bhattacharya \& van~den~Heuvel~(1991) and
Verbunt \& Lewin (this book) for a review.

\begin{figure}[b]
  \centering
    \centerline{\resizebox{8cm}{!}{\includegraphics{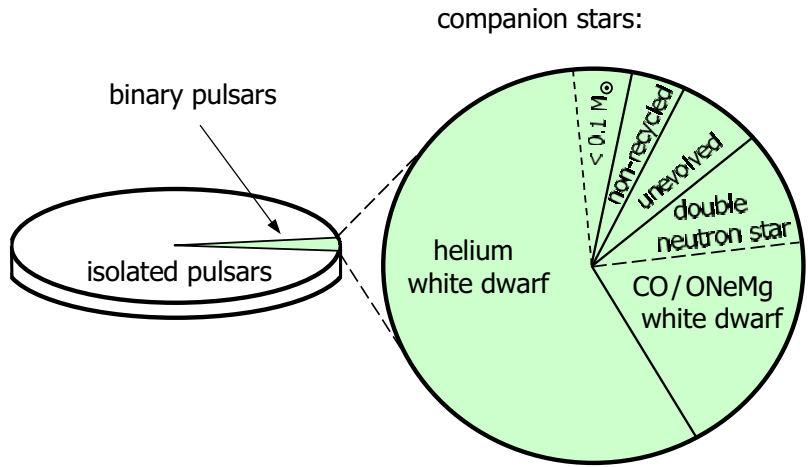}}}
  \caption{Illustration of the relative distribution of all $\sim\! 1500$ radio pulsars observed. 
           About 4$\,$\% are members of a binary system. The four main classes of binary pulsars are 
           defined according to their formation history -- see text.}
  \label{magnetosphere}
\end{figure}
\newpage
\setlength{\tabcolsep}{5.0pt}
   \begin{table*}[h]
     \caption{Main categories and types of binaries with compact objects}
\begin{footnotesize}
        \begin{tabular}{l}
           \hline\hline\noalign{\smallskip}
             {X}-RAY BINARIES\\
             \noalign{\smallskip}
             \cline{1-1}
        \end{tabular}
        \begin{tabular}{lllll}
           \noalign{\smallskip}
           \noalign{\smallskip}
           \noalign{\smallskip}
           \noalign{\smallskip}
             main type & & sub-type & & obs. example\\
           \noalign{\smallskip}
           \hline
           \noalign{\smallskip}
                 high-mass donor & & `standard' HMXB & & Cen X--3, P$_{\rm orb}=2.087^{\rm d}$ (NS)\\
                 $(M_{\rm donor} \ge 10\,M_\odot)$ & & & & Cyg X--1, P$_{\rm orb}=5.60^{\rm d}$ (BH)\\
                                 & & wide-orbit HMXB & & X Per, P$_{\rm orb}=250^{\rm d}$ (NS)\\
           \noalign{\smallskip}
           \cline{3-5}
           \noalign{\smallskip}
                            & & Be-star HMXB & & A0535+26, P$_{\rm orb}=104^{\rm d}$ (NS)\\
           \noalign{\smallskip}
           \cline{1-5}
           \noalign{\smallskip}
                             low-mass donor & & Galactic disk LMXB & & Sco X--1, P$_{\rm orb}=0.86^{\rm d}$ (NS)\\
                             $(M_{\rm donor} \le 1\,M_\odot)$ & & soft {X}-ray transient & & A0620--00, P$_{\rm orb}=7.75^{\rm hr}$ (BH)\\
                             & & globular cluster & & X 1820--30, P$_{\rm orb}=11^{\rm min}$ (NS)\\
           \noalign{\smallskip}
           \cline{3-5}
           \noalign{\smallskip}
                             & & millisecond X-ray pulsar & & SAX J1808.4--36, P$_{\rm orb}=2.0^{\rm d}$ (NS)\\
           \noalign{\smallskip}
           \cline{1-5}
           \noalign{\smallskip}
                             intermediate-mass donor& & & & Her X--1, P$_{\rm orb}=1.7^{\rm d}$ (NS)\\
                             $(1 < M_{\rm donor}/M_\odot < 10)$ & & & & Cyg X--2, P$_{\rm orb}=9.8^{\rm d}$ (NS)\\
                                                                & & & & V 404 Cyg, P$_{\rm orb}=6.5^{\rm d}$ (BH)\\
        \end{tabular}
        \bigskip
        \begin{tabular}{l}
           \hline\hline\noalign{\smallskip}
             BINARY RADIO PULSARS\\
             \noalign{\smallskip}
             \cline{1-1}
        \end{tabular}
        \begin{tabular}{lllll}
             main type & & sub-type & & obs. example\\
           \noalign{\smallskip}
           \hline
           \noalign{\smallskip}
                                   `high-mass' companion & & NS + NS (double) & & PSR 1913+16, P$_{\rm orb}=7.75^{\rm hr}$\\
                  $(0.5 \le M_{\rm c}/M_\odot \le 1.4)$ & & NS + (ONeMg) WD & & PSR 1435--6100, P$_{\rm orb}=1.35^{\rm d}$\\
                                 & & NS + (CO) WD & & PSR 2145--0750, P$_{\rm orb}=6.84^{\rm d}$\\
           \noalign{\smallskip}
           \cline{1-5}
                                 `low-mass' companion & & NS + (He) WD & & PSR 0437--4715, P$_{\rm orb}=5.74^{\rm d}$\\
                                 $(M_{\rm c} < 0.45\,M_\odot$) & & & & PSR 1640+2224, P$_{\rm orb}=175^{\rm d}$\\
           \noalign{\smallskip}
           \cline{1-5}
           \noalign{\smallskip}
                                 non-recycled pulsar & & (CO) WD + NS & & PSR 2303+46, P$_{\rm orb}=12.3^{\rm d}$\\
           \noalign{\smallskip}
           \cline{1-5}
           \noalign{\smallskip}
            unevolved companion & & B-type companion & & PSR 1259--63, P$_{\rm orb}=3.4^{\rm yr}$\\
                                & & low-mass companion & & PSR 1820--11, P$_{\rm orb}=357^{\rm d}$\\
        \end{tabular}
        \bigskip
        \begin{tabular}{l}
           \hline\hline\noalign{\smallskip}
             CV-LIKE BINARIES\\
             \noalign{\smallskip}
             \cline{1-1}
        \end{tabular}
        \begin{tabular}{lllll}
             main type & & sub-type & & obs. example\\
           \noalign{\smallskip}
           \hline
           \noalign{\smallskip}
                             novae-like systems & & $(M_{\rm donor}\le M_{\rm WD})$ & & DQ Her, P$_{\rm orb}=4.7^{\rm hr}$\\
                             & & & &                                                 SS Cyg, P$_{\rm orb}=6.6^{\rm hr}$\\
           \noalign{\smallskip}
           \cline{3-5}
           \noalign{\smallskip}
                      super soft {X}-ray sources & & $(M_{\rm donor}>M_{\rm WD})$ & & CAL 83, P$_{\rm orb}=1.04^{\rm d}$\\
                      & & & &                                                         CAL 87, P$_{\rm orb}=10.6^{\rm hr}$\\
           \noalign{\smallskip}
           \cline{1-5}
           \noalign{\smallskip}
            AM CVn systems (RLO) & & (CO) WD + (He) WD & & AM CVn, P$_{\rm orb}=22^{\rm min}$\\
           \noalign{\smallskip}
           \cline{3-5}
           \noalign{\smallskip}
            double WD (no RLO) & & (CO) WD + (CO) WD & & WD1204+450, P$_{\rm orb}=1.6^{\rm d}$\\
           \noalign{\smallskip}
           \cline{1-5}
           \noalign{\smallskip}
            sdB-star systems  & & (sdB) He-star + WD & & KPD 0422+5421, P$_{\rm orb}=2.16^{\rm hr}$\\
           \noalign{\smallskip}
           \hline\hline
        \end{tabular}
        \label{binaries}
\end{footnotesize}
   \end{table*}
\setlength{\tabcolsep}{6pt}

\newpage
\section{Binary stellar evolution and final compact objects}
In order to understand how neutron stars, black holes and white dwarfs can be formed in binary systems,
a brief overview of the basic elements of the evolution of single stars is necessary.
We refer to e.g. Cox \& Giuli~(1968) and Kippenhahn \& Weigert~(1990) for further details.

\subsection{Summary of the evolution of single stars}
The evolution of a star is driven by a rather curious property of a self-gravitating gas in
hydrostatic equilibrium, described by the virial theorem, namely that the radiative loss of energy
of such a gas causes it to contract and herewith, due to release of gravitational potential energy, 
to increase its temperature.
Thus, while the star tries to cool itself by radiating away energy from its surface,
it gets hotter instead of cooler (i.e. it has a `negative heat capacity'). The more it radiates
to cool itself, the more it will contract, the hotter it gets and the more it is forced 
to go on radiating. Clearly, this `vicious virial cycle' is an unstable situation in the long run
and explains why the star, starting out as an interstellar gas globe, must
finally end its life as a compact object. 
In the meantime the star spends a
considerable amount of time in the intermediate stages, which are called: `the main-sequence',
`the giant branch' etc. It is important to realize that stars do {\em not} shine because they
are burning nuclear fuel. They shine because they are hot due to their history of
gravitational contraction.\\
A massive star ($M \ga 10\,M_{\odot}$) evolves through cycles of nuclear burning
alternating with stages of exhaustion of nuclear fuel in the stellar core until its core
is made of iron, at which point further fusion requires, rather than releases, energy.
The core mass of such a star becomes larger than the Chandrasekhar limit, the maximum mass possible
for an electron-degenerate configuration ($\sim\!1.4\,M_{\odot}$). Therefore the core
implodes to form a neutron star or black hole. The gravitational energy
released in this implosion ($4\times 10^{53}$~erg $\simeq 0.15\,M_{\rm core}c^2$) is far more than
the binding energy of the stellar envelope, causing the collapsing star to violently explode and eject
the outer layers of the star, with a speed of $\sim\!10^4$~km$\,$s$^{-1}$, in a supernova event. 
The final stages during and beyond carbon burning are very short lasting ($\sim\!60$~yr for
a $25\,M_{\odot}$ star) because most of the nuclear energy generated in the interior
is liberated in the form of neutrinos which freely escape without interaction with the
stellar gas and thereby lowering the outward pressure and accelerating the contraction
and~nuclear~burning.\\
 \begin{table*}[t]
  \caption{End products of stellar evolution as a function of initial mass}
    \begin{tabular}{llll}
     \hline \hline
     & & \multicolumn{2}{c} {Final product}\\
     \noalign{\smallskip}
     \cline{3-4} 
     \noalign{\smallskip}
     Initial mass & He-core mass & Single star & Binary star\\
     \hline
     $< 2.3\,M_{\odot}$ & $< 0.45\,M_{\odot}$ & CO white dwarf & He white dwarf\\
     $2.3 - 6\,M_{\odot}$ & $0.5 - 1.9\,M_{\odot}$ & CO white dwarf & CO white dwarf\\
     $6 - 8\,M_{\odot}$ & $1.9 - 2.1\,M_{\odot}$ & O-Ne-Mg white dwarf & O-Ne-Mg white dwarf\\
      & & or C-deflagration SN ? & \\
     $8 - 12\,M_{\odot}$ & $2.1 - 2.8\,M_{\odot}$ & neutron star & O-Ne-Mg white dwarf\\
     $12 - 25\,M_{\odot}$ & $2.8 - 8\,M_{\odot}$ & neutron star & neutron star\\
     $> 25\,M_{\odot}$ & $>8\,M_{\odot}$ & black hole & black hole\\
     \hline \hline
    \end{tabular}
  \label{final_star}
 \end{table*}
Less massive stars ($M < 8\,M_{\odot}$) suffer from the occurrence of degeneracy in the core
at a certain point of evolution. Since for a degenerate gas the pressure only depends on
density and not on the temperature, there will be no stabilizing expansion and subsequent
cooling after the ignition. Hence, the sudden temperature rise
(due to the liberation of energy after ignition) causes a run-away nuclear energy generation
producing a so-called `flash'. In stars with $M < 2.3\,M_{\odot}$ the helium core becomes
degenerate during hydrogen shell burning and, when its core mass $M_{\rm He}$ reaches $0.45\,M_{\odot}$, helium
ignites with a flash. The helium flash is, however, not violent enough to disrupt the star. 
Stars with masses in the range $2.3 < M/M_{\odot} < 8$ ignite carbon with a flash.
Such a carbon flash was believed to perhaps disrupt the entire star in a so-called carbon-deflagration
supernova. However, recent observations of white dwarfs in Galactic clusters that
still contain stars as massive as $8^{+3}_{-2}\,M_{\odot}$ (Reimers \& Koester~1988; Weidemann 1990) indicate
that such massive stars still terminate their life as a white dwarf. They apparently
shed their envelopes in the AGB-phase before carbon ignites violently. Furthermore,
stars in close binary systems, which are the prime objects in this review, will
have lost their envelope as a result of mass transfer via Roche-lobe overflow. This is also the reason
why in binary systems the lower ZAMS mass-limit for producing a neutron star is
somewhat larger than for an isolated star.

The possible end-products and corresponding initial masses are listed in Table~\ref{final_star}.
It should be noted that the actual values of the different mass ranges are only known
approximately due to considerable uncertainty in our knowledge of the evolution of
massive stars. Prime causes of this uncertainty include limited understanding of
the mass loss undergone by stars in their various evolutionary stages (see Sect.~\ref{heliumevol}).
To make a black hole, the initial ZAMS stellar mass must exceed at least $20\,M_{\odot}$ (Fryer 1999),
or possibly, $25\,M_{\odot}$. According to MacFadyen, Woosley \& Heger~(2001), stars $>40\,M_{\odot}$ 
form black holes directly (collapsars type~I) whereas stars
in the interval $25 < M/M_{\odot} < 40$ produce black holes after a `failed supernova explosion' (collapsars type~II). 
From an analysis of black hole binaries it seems that a mass-fraction of $\sim\!0.35$ must have been
ejected in the (symmetric) stellar core collapse leading to the formation of a black~hole 
(Nelemans, Tauris \& van~den~Heuvel 1999). 
Another fundamental problem is understanding convection, in particular in stars that
consist of layers with very different chemical composition. Finally, there is the unsolved question
of whether or not the velocity of convective gas cells may carry them beyond the boundary
of the region of the star which is convective according to the Schwarzschild criterion.
For example, inclusion of this so-called overshooting in evolutionary calculations
decreases the lower mass-limit for neutron star progenitors.

\subsubsection{Three timescales of stellar evolution}
There are three fundamental timescales of stellar evolution. When the hydrostatic equilibrium
of a star is disturbed (e.g. because of sudden mass loss), the star will restore this equilibrium on
a so-called dynamical (or pulsational) timescale:
\begin{equation}
  \tau _{\rm dyn} = \sqrt{R^3/GM} \qquad\simeq 30\;{\rm min}\; (R/R_{\odot})^{3/2}\;(M/M_{\odot})^{-1/2} 
\end{equation}
When the thermal equilibrium of a star is disturbed, it will restore this equilibrium on
a thermal (or Kelvin-Helmholtz) timescale, which is the time it takes to emit all of its
thermal energy content at its present luminosity:
\begin{equation}
  \tau _{\rm th} = GM^2/RL \qquad\simeq 30\;{\rm Myr}\; (M/M_{\odot})^{-2}
\end{equation}
The third stellar timescale is the nuclear one, which is the time needed for the star to exhaust
its nuclear fuel reserve (which is proportional to $M$), at its present fuel consumption rate
(which is proportional to $L$), so this timescale is given by:
\begin{equation}
  \tau _{\rm nuc} \simeq 10\;{\rm Gyr}\; (M/M_{\odot})^{-2.5}
\end{equation}
In calculating the above mentioned timescales we have assumed a mass-luminosity relation: $L \propto M^{3.5}$ and a
mass-radius relation for main-sequence stars: $R \propto M^{0.5}$. Both of these relations are fairly good approximations for
$M \ge M_{\odot}$. Hence, it should also be noted that the rough numerical estimates of these timescales only apply to ZAMS stars.

\subsection{The variation of the outer radius during stellar evolution}
\begin{figure}
  \centering
    \centerline{\resizebox{10.5cm}{!}{\includegraphics{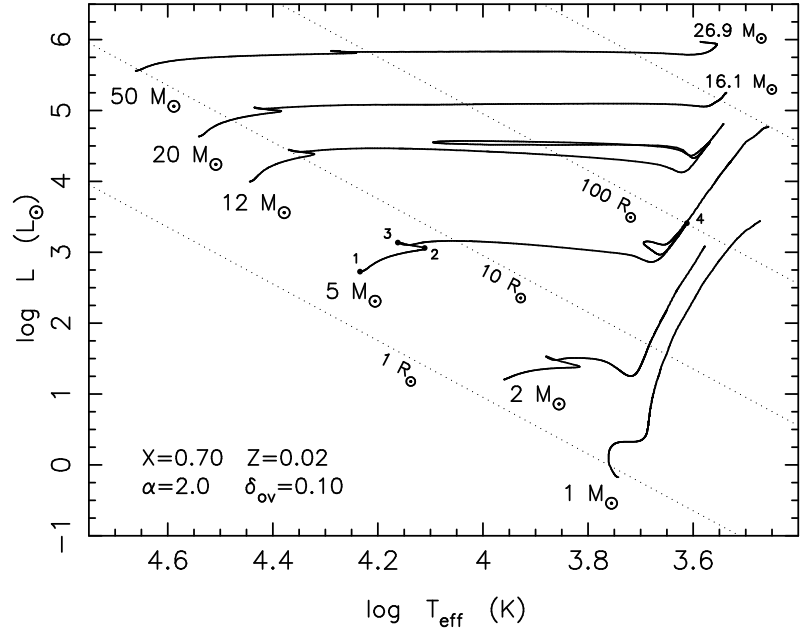}}}
   \caption{Stellar evolutionary tracks in the HR-diagram.}
  \label{HR-diagram}
\end{figure}

\begin{figure}
  \centering
    \centerline{\resizebox{10.5cm}{!}{\includegraphics{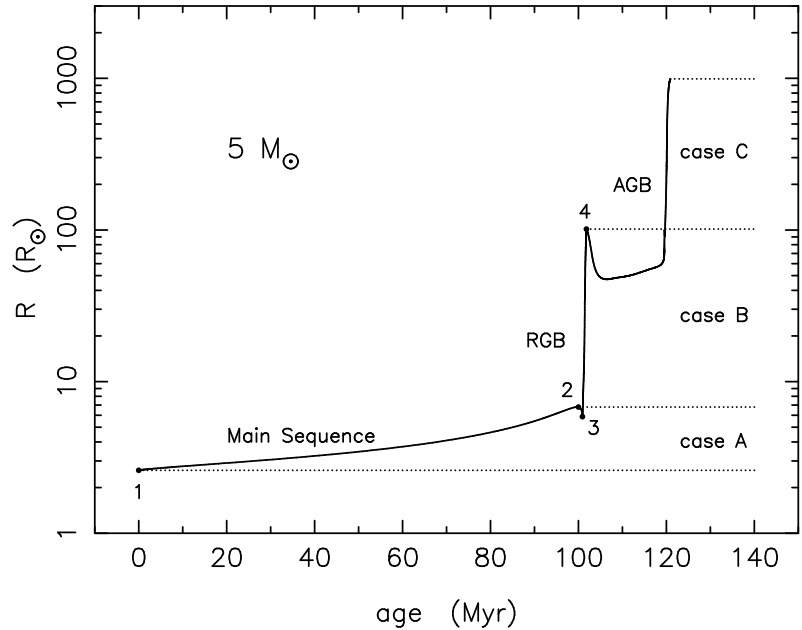}}}
  \caption{Evolutionary change of the radius of the $5\,M_{\odot}$ star plotted in the figure above
           (Fig.~\ref{HR-diagram}).
           The ranges of radii for mass transfer to a companion star in a binary system
           according to RLO cases~A, B and C are indicated -- see Sect.~\ref{RLO} for an explanation.}
  \label{R-evol}
\end{figure}

Figure~\ref{HR-diagram} depicts the evolutionary tracks in the Hertzsprung-Russel diagram of six
different stars ($50\,M_{\odot}$, $20\,M_{\odot}$, $12\,M_{\odot}$, $5\,M_{\odot}$, $2\,M_{\odot}$ and $1\,M_{\odot}$).
We calculated these tracks using Eggleton's evolutionary code (e.g. Pols~et~al.~1995,$\,$1998).
The observable stellar parameters are: luminosity ($L$), radius ($R$) and effective surface
temperature ($T_{\rm eff}$). Their well-known relationship is given by: $L=4\pi R^2\sigma T_{\rm eff}^4$. 
In Figure~\ref{R-evol} we have plotted our calculation of stellar radius as a function of age for the $5\,M_{\odot}$ star.
Important evolutionary stages are indicated in the figures. Between points~1 and 2 the star is in the
long-lasting phase of core hydrogen burning (nuclear timescale). At point~3 hydrogen ignites in a shell
around the helium core. For stars more massive than $1.2\,M_{\odot}$ the entire star briefly contracts
between points~2 and 3, causing its central temperature to rise. 
When the central temperatures reaches $T\sim\!10^8$~K, core helium ignites (point~4). At this moment the star has 
become a red giant, with a dense core and a very large radius. During helium burning it describes a loop
in the HR-diagram. Stars with $M\ge 2.3\,M_{\odot}$ move from points~2 to 4 on a thermal timescale and
describe the helium-burning loop on a (helium) nuclear timescale following point~4. 
Finally, during helium shell burning the outer
radius expands again and at carbon ignition the star has become a red supergiant on the asymptotic giant branch (AGB).\\
The evolution of less massive stars ($M < 2.3\,M_{\odot}$) takes a somewhat different course.
After hydrogen shell ignition the helium core becomes degenerate and the hydrogen burning shell generates
the entire stellar luminosity. While its core mass grows, the star gradually climbs upwards along the red
giant branch until it reaches helium ignition with a flash. For all stars less massive than about
$2.3\,M_{\odot}$ the helium core has a mass of about $0.45\,M_{\odot}$ at helium flash ignition.
The evolution described above depends only slightly on the initial chemical composition and effects of convective
overshooting.

\subsubsection{The core mass -- radius relation for low-mass RGB stars}
For a low-mass star ($\la 2.3\,M_{\odot}$) on the red giant branch (RGB) the growth in core mass is directly 
related to its luminosity, as this luminosity is entirely generated by hydrogen shell burning.
As such a star, composed of a small dense core surrounded by an extended convective envelope,
is forced to move up the Hayashi track its luminosity increases strongly with only a fairly modest
decrease in temperature. Hence one also finds a relationship between the giant's radius and the mass of
its degenerate helium core -- almost entirely independent of the mass present in the hydrogen-rich
envelope (Refsdal \& Weigert 1971; Webbink, Rappaport \& Savonije 1983). This relationship is very important for
LMXBs and wide-orbit binary pulsars since, as we shall see later on, it results in a relationship
between orbital period and white dwarf mass.

\subsection{The evolution of helium stars}
\label{heliumevol}
For low-mass stars, the evolution of the helium core in post main-sequence stars is practically
independent of the presence of an extended hydrogen-rich envelope. However, for more massive
stars ($> 2.3\,M_{\odot}$) the evolution of the core of an isolated star differs from that of a
naked helium star (i.e. a star which has lost
its hydrogen envelope via mass transfer in a close binary system). 
Thus, it is very important to study the giant phases of helium star evolution.
Pioneering studies in this field
are those of Paczy\'nski~(1971), Nomoto~(1984) and Habets~(1986). 
\begin{figure}[t]
  \centering
    \centerline{\resizebox{7cm}{!}{\includegraphics{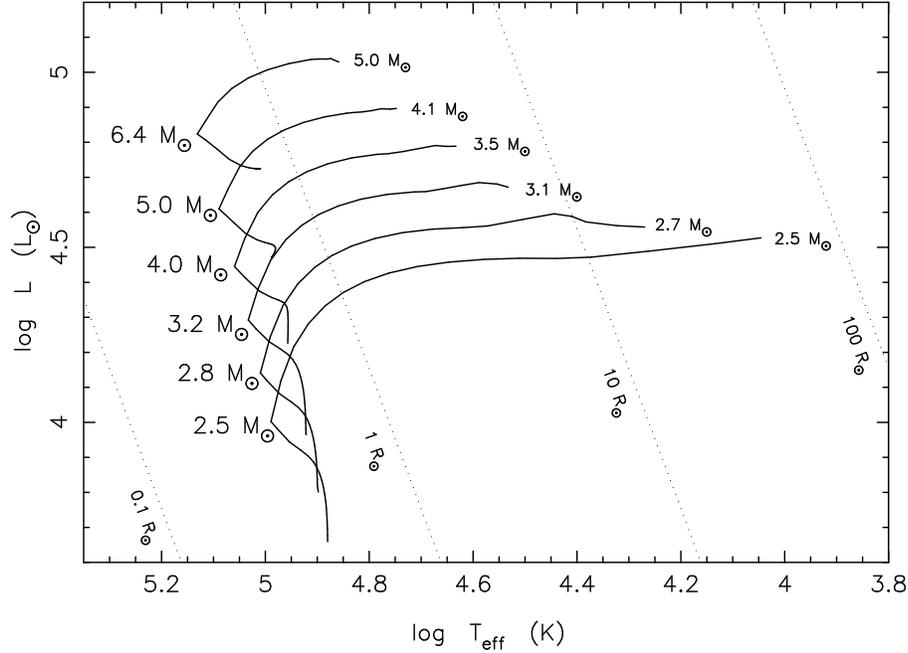}}}
  \caption{Evolutionary tracks of $2.5\,M_{\odot}-6.4\,M_{\odot}$ helium stars ({Y}=0.98, {Z}=0.02). The final stellar mass (after 
wind mass loss) is written at the end of the tracks. The expansion of low-mass helium stars in close binaries often
results in a second mass-transfer phase (case~BB RLO). This plot was made with data provided by O.~Pols~(2002, private communication).} 
  \label{habets_radii}
\end{figure}
\begin{figure}[t]
  \centering
    \centerline{\resizebox{12cm}{!}{\includegraphics{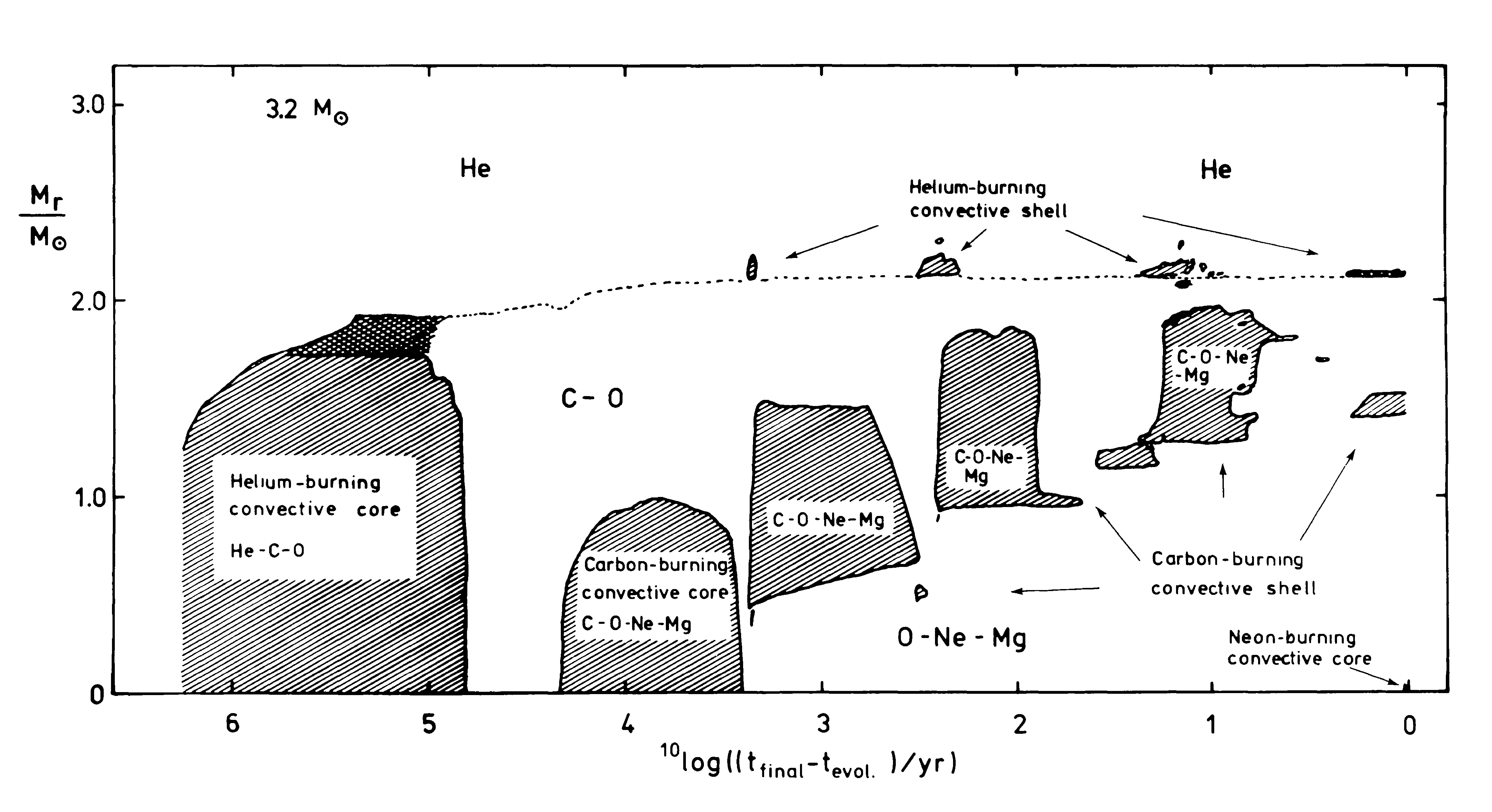}}}
  \caption{The interior evolution of a $3.2\,M_{\odot}$ helium star. Hatched regions are
           convective; double hatched regions are semi-convective. The broken line indicates the region of
           maximum net energy generation in the helium-burning shell (approx. the boundary between the
           {CO} core and its adjacent helium layer). After Habets~(1986).}
  \label{habets_interior}
\end{figure}
$\hspace{-0.2cm}$Of particular interest are the low-mass helium stars ($M_{\rm He} < 3.5\,M_{\odot}$) since they
swell up to large radii during their late evolution  -- see Fig.~\ref{habets_radii}.
This may cause an additional phase of mass transfer from the naked helium star to its companion
(often referred to as so-called case~BB mass transfer). Recent detailed studies of helium stars
in binaries have been performed by Dewi~et~al.~(2002). Using helium star models ({Z}=0.03, {Y}=0.97)
calculated by O.~Pols~(2002, private~communication), we fitted the helium star ZAMS radii as a function of mass:
\begin{equation}
  R_{\rm He} = 0.212\;(M_{\rm He}/M_{\odot})^{0.654} \;\; R_{\odot}
\end{equation}
It is important to realize that helium cores in binaries have tiny envelopes of hydrogen ($<0.01\,M_{\odot}$)
when they detach from RLO. This has important effects on their subsequent radial evolution (e.g. Han~et~al.~2002).\\
The evolution of more massive helium stars (Wolf-Rayet stars) is also quite important. There is currently
not a clear agreement on the rate of intense wind mass-loss from Wolf-Rayet stars (e.g. Wellstein \& Langer~1999;
Nugis \& Lamers~2000; Nelemans \& van~den~Heuvel~2001). 
A best-estimate fit to the wind mass-loss rate of Wolf-Rayet stars is, for example, given by Dewi~et~al.~(2002):
\begin{equation}
  \dot{M}_{\rm He,\;wind} = \, \left\{ \begin{array}{lll}
   2.8\times 10^{-13}\,(L/L_{\odot})^{1.5}  & \; M_{\odot}\,\mbox{yr}^{-1}, & \quad\log\,(L/L_{\odot}) \ge 4.5\\ 
   4.0\times 10^{-37}\,(L/L_{\odot})^{6.8}  & \; M_{\odot}\,\mbox{yr}^{-1}, & \quad\log\,(L/L_{\odot}) < 4.5\\ 
               \end{array}
         \right.
\end{equation}
The uncertainty in determining this rate also effects our knowledge of the threshold mass for core collapse 
into a black hole (Schaller~et~al. 1992; Woosley, Langer \& Weaver 1995; Brown, Lee \& Bethe 1999).
Very important in this respect is the question whether the helium star is ``naked'' or ``embedded'' -- i.e.
is the helium core of the massive star surrounded by a thick hydrogen mantle$\,$? In the latter case this helium `star'
does not lose much mass in the form of a wind and it can go through all burning stages and terminate as a black hole.
Single star evolutionary models suggest that this happens above an initial stellar mass $\ge19\,M_{\odot}$,
as around this mass a sudden increase in the mass of the collapsing iron core occurs to $\ge 1.9\,M_{\odot}$ 
(Woosley \& Weaver 1995). In order to form a black hole in a close binary, it is best to keep the helium core
embedded in its hydrogen envelope as long as possible, i.e. to start from a wide ``case~C'' binary evolution,
as has been convincingly argued by Brown, Lee \& Bethe~(1999); Wellstein \& Langer~(1999); Brown et~al.~(2001)
and Nelemans \& van~den~Heuvel~(2001). In this case common envelope evolution (Sect.~\ref{CE}) leads to a narrow system
consisting of the evolved helium core and the low-mass companion. The helium core collapses to a black hole
and produces a supernova in this process, shortly after the spiral-in. When the low-mass companion evolves
to fill its Roche-lobe these systems are observed as soft {x}-ray transients (SXTs) -- see Sect.~\ref{SXT}.

\newpage
\section{Roche-lobe overflow (RLO) -- cases A, B and C}
\label{RLO}
\begin{figure}
  \begin{center}
    \centerline{\resizebox{10.5cm}{!}{\includegraphics{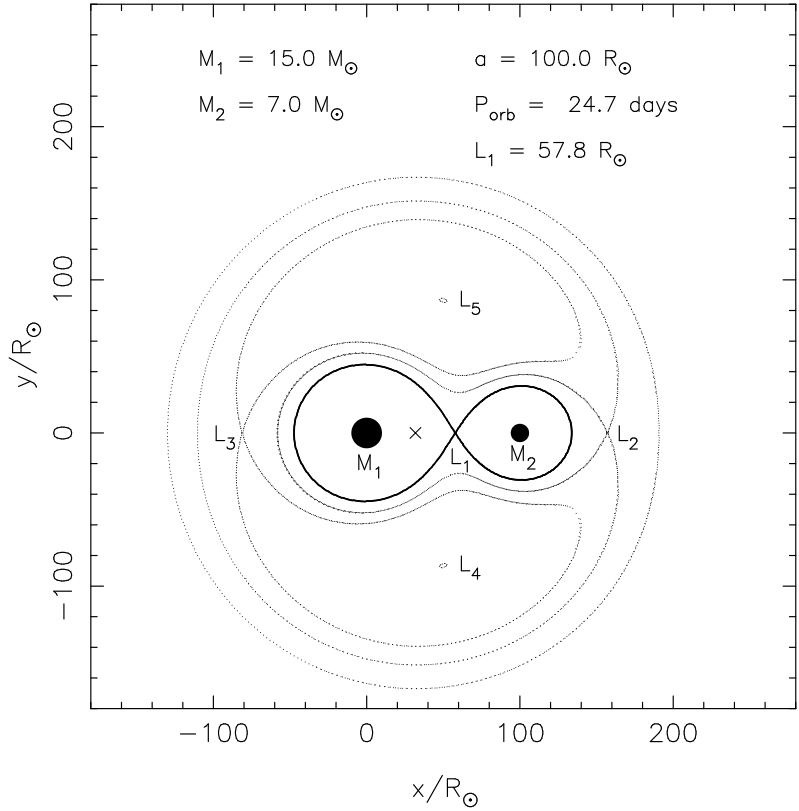}}}
    \caption{A cross-section in the equatorial plane of the critical equipotential surfaces in a binary. 
             The thick curve crossing through $L_1$ is the Roche-lobe.}
    \label{RLO_fig}
  \end{center}
\end{figure}

The effective gravitational potential in a binary system is determined by the masses of the stars and
the centrifugal force arising from the motion of the two stars around one another.
One may write this potential as:
\begin{equation}
  \Phi = -\frac{GM_1}{r_1} -\frac{GM_2}{r_2} - \frac{\Omega ^2 r_3^2}{2}
\end{equation}
where $r_1$ and $r_2$ are the distances to the center of the stars with mass $M_1$ and $M_2$,
respectively; $\Omega$ is the orbital angular velocity; and $r_3$ is the distance to the
rotational axis of the binary. It is assumed that the stars are small with respect
to the distance between them and that they revolve in circular orbits, i.e. $\Omega = \sqrt{GM/a^3}$.
In a binary where tidal forces have circularized the orbit, and brought the two stellar components
into synchronized co-rotation, one can define fixed equipotential surfaces in a co-moving frame
(see e.g. van~den~Heuvel 1994). 
The equipotential surface passing through the first Lagrangian point, $L_1$ defines
the `pear-shaped' Roche-lobe -- see the cross-section in Fig.~\ref{RLO_fig}.
If the initially more massive star (the donor) evolves to fill its Roche-lobe the unbalanced pressure
at $L_1$ will initiate mass transfer (Roche-lobe overflow,~RLO) onto its companion star
(the accretor). 
The radius of the donor's Roche-lobe, $R_L$ is defined as that of a sphere with the same volume as
the lobe. It is a function only of the orbital separation,~$a$ and the mass ratio, 
$q\equiv M_{\rm donor}/ M_{\rm accretor}$ of the binary components. It can be approximated as
(Eggleton~1983):
\begin{equation}
    \frac{R_L}{a} = \frac{0.49\,q^{2/3}}{0.6\,q^{2/3} + \ln(1+q^{1/3})}
\label{Eggleton}
\end{equation}
A star born in a close binary system with a radius smaller than that of its Roche-lobe may,
either because of expansion of its envelope at a later evolutionary stage or because the binary
shrinks sufficiently as a result of orbital angular momentum losses, begin RLO. 
The further evolution of the system will now depend on the evolutionary state and structure
of the donor star at the onset of the overflow, which is determined by $M_{\rm donor}$ and $a$,
as well as the nature of the accreting star.
Kippenhahn \& Weigert~(1967) define three types of RLO: cases~A, B and C.
In case~A, the system is so close that the donor star begins to fill its Roche-lobe during 
core-hydrogen burning; in case~B the primary star begins to fill its Roche-lobe after the end
of core-hydrogen burning but before helium ignition; in case~C it overflows its Roche-lobe
during helium shell burning or beyond. It is clear from Fig.~\ref{R-evol} that cases~B and C occur over a
wide range of radii (orbital periods); case~C even up to orbital periods of $\sim\!10$~years. The precise orbital
period ranges for cases A, B and C depend on the initial donor star mass and on the mass ratio.
Once the RLO has started it continues until the donor has lost its hydrogen-rich envelope 
(typically $\ge$ 70\% of its total mass) and subsequently no longer fills its Roche-lobe.
 
\subsection{The orbital angular momentum balance equation}
~\label{OAMB}
The orbital angular momentum of a binary system is given by:
\begin{equation}
  J_{\rm orb} = \frac{M_1 M_2}{M}\,\Omega \,a^2 \,\sqrt{1-e^2}
~\label{Jorb}
\end{equation}
where $a$ is the separation between the stellar components; $M_1$ and $M_2$ are the masses of
the accretor and donor star, respectively; $M=M_1 + M_2$ and the orbital angular velocity,
$\Omega = \sqrt{GM/a^3}$. Here $G$ is the constant of gravity.
As mentioned earlier, tidal effects acting on a near-RLO (giant) star will circularize the
orbit on a short timescale of $\sim\!10^4$~yr (Verbunt \& Phinney 1995).
In the following we therefore neglect any small eccentricity ($e=0$).
A simple logarithmic differentiation of the above equation yields the rate of change in orbital separation:
\begin{equation}
  \frac{\dot{a}}{a} = 2\,\frac{\dot{J}_{\rm orb}}{J_{\rm orb}} - 2\,\frac{\dot{M}_1}{M_1} 
                    - 2\,\frac{\dot{M}_2}{M_2} + \frac{\dot{M}_1 + \dot{M}_2}{M}
  \label{adot}
\end{equation}
where the total change in orbital angular momentum is given by:
\begin{equation}
  \frac{\dot{J}_{\rm orb}}{J_{\rm orb}} = \frac{\dot{J}_{\rm gwr}}{J_{\rm orb}}
       +\frac{\dot{J}_{\rm mb}}{J_{\rm orb}} + \frac{\dot{J}_{\rm ls}}{J_{\rm orb}}
       +\frac{\dot{J}_{\rm ml}}{J_{\rm orb}}                         
  \label{Jdot}
\end{equation}
These two equations constitute the orbital angular momentum balance equation and will now be discussed
in more detail.
The first term on the right-hand side of Eq.~\ref{Jdot} gives the change in orbital angular momentum
due to gravitational wave radiation (Landau \& Lifshitz~1958):
\begin{equation}
  \frac{\dot{J}_{\rm gwr}}{J_{\rm orb}} = - \frac{32\,G^3}{5\,c^5}\,\frac{M_1 M_2 M}{a^4} \qquad \rm{s}^{-1}
\end{equation}
where $c$ is the speed of light in vacuum. The validity of this mechanism has been beautifully
demonstrated in PSR~1913+16 which is an ideal GTR-laboratory (e.g. Taylor \& Weisberg 1989).
For sufficiently narrow orbits the above equation becomes the dominant term in Eq.~\ref{Jdot}
and will cause $a$ to decrease. Therefore, the orbits of very narrow binaries will tend to
continuously shrink, forcing the components into contact. Gravitational radiation losses are a
major force driving the mass transfer in very narrow binaries, such as CVs and LMXBs (Faulkner~1971). 

The second term in Eq.~\ref{Jdot} arises due to so-called magnetic braking.
The presence of magnetic stellar winds has long been known to decelerate the rotation of low-mass
stars (e.g. Kraft~1967; Skumanich~1972; Sonderblom~1983). The loss of spin angular momentum is
caused by the magnetic field which exists
as a result of chromospheric coronal activity of cool $\la 1.5\,M_{\odot}$ stars
with subphotospheric convection zones (Mestel~1984). In tight synchronized binaries, the loss
of spin angular momentum occurs at the expense of the orbital angular momentum. As a result
the orbital period decreases while the stellar components spin~up, due to tidal forces, and approach one another.
Based on Skumanich's observations, Verbunt \& Zwaan~(1981) derived an expression for the effect
of the magnetic braking and applied it to LMXBs by extrapolating, the dependence of the magnetic
braking on the orbital angular velocity, down to very short orbital periods (of the order $\sim\,$hours): 
\begin{equation}
   \frac{\dot{J}_{\rm mb}}{J_{\rm orb}} \simeq -0.5\times 10^{-28}\,f_{\rm mb}^{-2}\; 
              \frac{k^2 R_2^4}{a^5}\frac{G M^3}{M_1 M_2}  \qquad \rm{s}^{-1}
\end{equation}
(in cgs units) where $R_2$ is the radius of the mass-losing star; $k^2$ is its gyration radius and $f_{\rm mb}$
is a constant of order unity.
However, a fundamental law of angular momentum loss is unknown for rapidly rotating stars.
Rappaport, Verbunt \& Joss~(1983) investigated a weaker dependency on the stellar radius.
Meanwhile, it now seems that the necessary stellar activity may saturate for rotation periods
shorter than $\sim$ 2--3~days (e.g. Rucinski~1983; Vilhu \& Walter~1987) which leads to a much flatter dependence
of the angular momentum loss rate on the angular velocity ($\dot{J}_{\rm mb} \propto \Omega ^{1.2}$) than is given by the
Skumanich-law ($\dot{J}_{\rm mb} \propto \Omega ^3$). 
Based partly on observational work,
Stepien~(1995) derived a new magnetic braking law which smoothly matches the Skumanich-law
dependence for wide systems to the dependence obtained by Rucinski~(1983) for short
orbital period ($\la 3$ days) systems: 
\begin{equation}
  \frac{\dot{J}_{\rm mb}}{J_{\rm orb}} \simeq -1.90\times 10^{-16}\; 
              \frac{k^2 R_2^2}{a^2}\frac{M^2}{M_1 M_2}\,e^{-1.50\times 10^{-5}/\Omega} \qquad \rm{s}^{-1}
\end{equation}
The two formulas above represent a strong and a weak magnetic braking torque, respectively,
and their relative strength can be compared in e.g. Tauris~(2001). For a recent discussion 
see also Eggleton~(2001).
It should be noted that for many years it has been thought that the magnetic field has to be anchored 
in underlaying 
radiative layers of a star (Parker~1955). However, recent observations and calculations, e.g. by Dorch \& Nordlund~(2001), seem to
suggest that even fully convective stars still operate a significant magnetic field. This conclusion has 
important consequences for the explanation of the observed period gap in CVs (Spruit \& Ritter~1983).
We encourage further investigations on this topic.

The third term ($ \dot{J}_{\rm ls}/J_{\rm orb}$) on the right-hand side of Eq.~(\ref{Jdot}) 
describes possible exchange of angular momentum
between the orbit and the donor star due to its expansion or contraction. Tauris~(2001) calculated 
the pre-RLO spin-orbit couplings in LMXBs and demonstrated that the sole nuclear expansion of a (sub)giant
donor in a tight binary will lead to an orbital period {\em decrease} by $\sim\!10\,$\%, prior to the onset
of the RLO mass transfer, as a result of tidal interactions. This effect is most efficient
for binaries with $2 < P_{\rm orb} < 5$ days. In more narrow orbits the donor star does not
expand very much and for wide binaries the tidal torque is weak. However, when the effect of magnetic
braking is included in the calculations prior to RLO it will dominate the loss of orbital angular momentum if its
corresponding torque is relatively strong. The tidal torque in LMXBs can be
determined by considering the effect of turbulent viscosity in the convective envelope of the donor
on the equilibrium tide (Terquem~et~al. 1998). In very wide orbit LMXBs
($P_{\rm orb} > 100$~days) the orbital separation will always {\em widen} prior to RLO since the 
stellar wind mass loss becomes very important for such giant stars.
To quantize this effect one can apply the Reimers'~(1975) wind mass-loss rate:
\begin{equation}
  \dot{M}_{\rm wind} = -4\times 10^{13} \,\eta _{RW} L\,R/M \qquad M_{\odot}\, \rm{yr}^{-1}
\end{equation}
where the luminosity, radius and mass of the mass-losing star are in solar units and $\eta _{RW}\simeq 0.5$ is a mass-loss parameter.\\
Spin-orbit couplings in {X}-ray binaries can also help to stabilize the
mass transfer processes in IMXBs with radiative donor stars (Tauris \& Savonije 2001). In such systems
the effect of pumping spin angular momentum into the orbit is clearly seen in the calculations as a 
result of a contracting mass-losing star in a tidally locked system. This
causes the orbit to widen (or shrink less) and survive the, otherwise dynamically unstable, mass transfer.\\
The tidal effects in eccentric high-mass binary systems are discussed in e.g.
Witte \& Savonije~(1999) and Witte~(2001). 
For massive donor stars ($> 8\,M_{\odot}$) one can use the mass-loss rates e.g. by de~Jager, Nieuwenhuijzen
\& van~der~Hucht~(1988). 

Finally, the last term on the right-hand side of Eq.~(\ref{Jdot}) represents the change in orbital angular momentum
caused by mass loss from the binary system. This is usually the dominant term in the orbital angular momentum
balance equation and its total effect is given by:
\begin{equation}
  \frac{\dot{J}_{\rm ml}}{J_{\rm orb}} = \frac{\alpha + \beta q^2 + \delta\gamma(1+q)^2}{1+q}\,
                                         \frac{\dot{M}_2}{M_2}
\end{equation}
where $\alpha$, $\beta$ and $\delta$ are the fractions of mass lost from the donor in the form
of a direct fast wind, the mass ejected from the vicinity of the accretor and from a circumbinary
coplanar toroid (with radius, $a_r = \gamma ^2 a$), respectively -- see van~den~Heuvel~(1994) and
Soberman, Phinney \&  van~den~Heuvel~(1997). The accretion efficiency of the accreting star is thus
given by: $\epsilon = 1 -\alpha -\beta -\delta$, or equivalently:
\begin{equation}
  \partial M_1 = -(1 -\alpha -\beta -\delta)\, \partial M_2
\label{epsilon}
\end{equation}
where $\partial M_2 < 0$ ($M_2$ refers to the donor star). 
These factors will be functions of time as the binary system evolves during the mass-transfer phase.

The general solution for calculating the change in orbital separation during the {X}-ray phase
is found by integration of the orbital angular momentum balance equation (Eq.~\ref{adot}). 
It is often a good approximation to assume $\dot{J}_{\rm gwr}, \dot{J}_{\rm mb} \ll \dot{J}_{\rm ml}$
during short RLO, and if $\alpha, \beta$ and $\delta$ are constant in time: 
\begin{equation}
  \frac{a}{a_0} = \Gamma _{ls}\,\displaystyle \left( \frac{q}{q_0}\right) ^{2\,(\alpha + \gamma\delta -1)}\;
                \left(\frac{q+1}{q_0+1}\right) ^{\textstyle\frac{-\alpha - \beta +\delta}{1-\epsilon}}\;
                \left(\frac{\epsilon q+1}{\epsilon q_0+1}\right) ^{3+2\,\textstyle\frac{\alpha \epsilon ^2 + \beta +
                      \gamma\delta (1-\epsilon )^2} {\epsilon (1-\epsilon)}}
\label{aa0}
\end{equation}
where the subscript `0' denotes initial values and $\Gamma _{ls}$ is factor of order unity 
to account for the tidal spin-orbit couplings ($\dot{J}_{\rm ls}$) other than the magnetic braking
($\dot{J}_{\rm mb}$). We remind the reader that $q\equiv M_{\rm donor}/M_{\rm accretor}$.

\subsection{Stability criteria for mass transfer}
The stability and nature of the mass transfer is very important in binary stellar evolution.
It depends on the response of the mass-losing donor star and of the Roche-lobe 
-- see Soberman, Phinney \& van~den~Heuvel~(1997) for a review.
If the mass transfer proceeds on a short timescale (thermal or dynamical) the system is unlikely
to be observed during this short phase; whereas if the the mass transfer proceeds on a nuclear
timescale it is still able to sustain a high enough accretion rate onto the neutron star or black hole
for the system to be observed as an {X}-ray source for a long time. 

When the donor star fills its Roche-lobe, and is perturbed by removal of mass, it falls out of 
hydrostatic and thermal equilibrium. In the process of re-establishing equilibrium the star will
either grow or shrink -- first on a dynamical (sound crossing) timescale, and then on a slower
thermal (Kelvin-Helmholtz) timescale. But also the Roche-lobe changes in response to the
mass transfer/loss. As long as the donor star's Roche-lobe continues to enclose the star the
mass transfer is stable. Otherwise it is unstable and proceeds on a dynamical timescale.
Hence the question of stability is determined by a comparison of the exponents in power-law fits
of radius to mass, $R \sim M^{\zeta}$, for the donor star and the Roche-lobe respectively:
\begin{equation}
  \zeta _{\rm donor} \equiv \frac{\partial \ln R_2}{\partial \ln M_2} \quad \wedge \quad 
  \zeta _L \equiv \frac{\partial \ln R_L}{\partial \ln M_2}
  \label{zeta}
\end{equation}
where $R_2$ and $M_2$ refer to the mass-losing donor star. Given $R_2 = R_L$ (the condition at 
the onset of RLO) the initial stability criteria becomes:
\begin{equation}
  \zeta _L \le \zeta _{\rm donor}
\end{equation}
where $\zeta _{\rm donor}$ is the adiabatic or thermal (or somewhere in between) response of the
donor star to mass loss. Note, that the stability might change during the mass-transfer phase
so that initially stable systems become unstable, or vice versa, later in the evolution
(e.g. Kalogera \& Webbink 1996).
The radius of the donor is a function of time and mass and thus: 
\begin{equation}
  \dot{R}_2 = \frac {\partial R_2}{\partial t}\Bigg\vert _{M_2} + R_2\,\zeta_{\rm donor}\frac{\dot{M}_2}{M_2}
  \label{Rdot}
\end{equation}
\begin{equation}
  \dot{R}_L = \frac {\partial R_L}{\partial t}\Bigg\vert _{M_2} + R_L\,\zeta_L\frac{\dot{M}_2}{M_2}
  \label{RLdot}
\end{equation}
The second terms on the right-hand sides follow from Eq.~(\ref{zeta}); the first term of Eq.~(\ref{Rdot}) is due to expansion of
the donor star as a result of nuclear burning (e.g. shell hydrogen burning on the RGB) and the first
term in Eq.~(\ref{RLdot}) represents changes in $R_L$ which are not caused by mass transfer -- such as
orbital decay due to gravitational wave radiation and tidal spin-orbit couplings.
Tidal couplings act to synchronize the orbit whenever the rotation of the donor is perturbed (e.g. as
a result of magnetic braking or an increase in the moment of inertia while the donor expands).
The mass-loss rate of the donor can be found as a self-consistent solution to Eqs.~(\ref{Rdot}) and
(\ref{RLdot}) assuming $\dot{R}_2 = \dot{R}_L$ for stable mass transfer.

\subsection{Response of the Roche-lobe to mass transfer/loss}
\begin{figure}
  \centering
    \centerline{\resizebox{10.8cm}{!}{\includegraphics{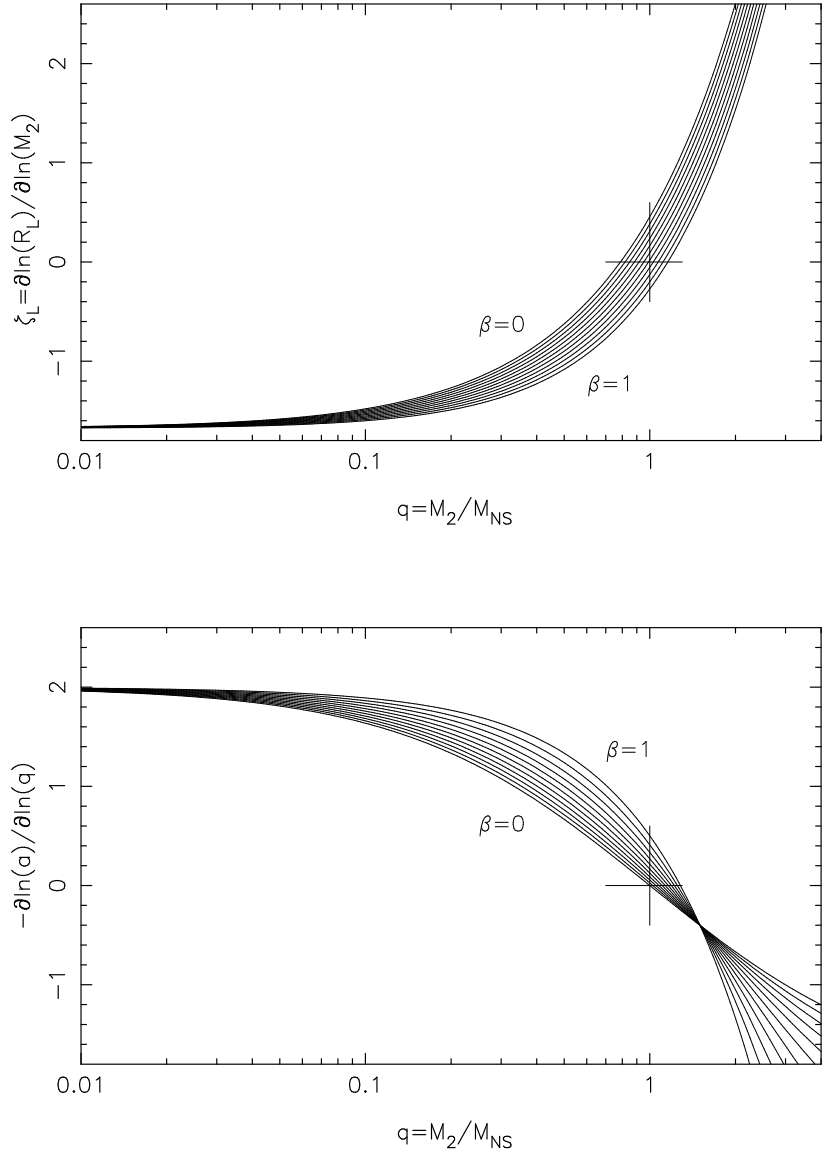}}}
  \caption{Top panel: the Roche-radius exponent ($R_L \propto M_2 ^{\zeta _L}$) for LMXBs as a
   function of $q$ and $\beta$. The different curves correspond to different constant values of $\beta$
   in steps of 0.1. Tidal effects are not taken into account and the mass loss is
   according to the isotropic re-emission model (i.e. $\alpha = 0, \delta =0$).
   A cross is shown to highlight the cases of $q=1$ and $\zeta _L = 0$. In the bottom panel we have
   plotted $-\partial \ln a/\partial \ln q$ as a function of $q$. 
   When this quantity is positive the orbit widens. This is the case when $q\le 1$.
   For more massive donor stars
   ($q>1$) the orbit shriks in response to mass transfer.
   Since $M_2$ (and hence $q$) is decreasing with time the evolution during the mass-transfer phase
   follows these curves from right to left, although $\beta$ need not be constant.
   After Tauris \& Savonije~(1999).}
  \label{zetafig}
\end{figure}

In order to study the dynamical evolution of an {X}-ray binary let us consider the cases where
tidal interactions and gravitational wave radiation can be neglected. We shall also assume that the
amount of mass lost directly from the donor star in the form of a fast wind,
or via a circumbinary toroid, is negligible compared
to the flow of material transfered via the Roche-lobe. Hence we have
$\dot{J}_{\rm gwr}=\dot{J}_{\rm mb}=\dot{J}_{\rm ls}=0$ and $\dot{J}_{\rm ml}/J_{\rm orb}=\beta q^2/(1+q)\times(\dot{M}_2/M_2)$.
This corresponds to the mode of `isotropic re-emission' where matter flows over from the donor star
onto the compact accretor (neutron star or black hole) in a conservative way, before a fraction, $\beta$ of
this material is ejected isotropically from the system with the specific angular momentum of the accretor,
i.e. $\dot{M}_1 = -(1-\beta)\dot{M}_2$, $dJ_{\rm orb} = (J_1/M_1)\,\beta dM_2$ and $J_1 = (M_2/M)\,J_{\rm orb}$.
In the above formalism this corresponds to $\alpha = \delta = 0$ and $\Gamma _{ls}=1$.
This is actually a good approximation for many real systems (with orbital periods larger than a few days).
Following Tauris \& Savonije~(1999) one can then combine Eqs.~(\ref{Eggleton}), (\ref{epsilon}) and (\ref{aa0})
and obtain an analytical  expression for $\zeta _L$:
\begin{eqnarray}
  \zeta _L & = & \frac{\partial \ln R_L}{\partial \ln M_2} = \left( \frac{\partial \ln a}{\partial \ln q} +
             \frac{\partial \ln (R_L/a)}{\partial \ln q} \right) \, \frac{\partial \ln q}{\partial \ln M_2}\nonumber \\
           & = & [1+(1-\beta )q]\psi \, + \, (5-3\beta )q
\end{eqnarray}
where
\begin{equation}
   \psi = \left[ -\frac{4}{3} -\frac{q}{1+q} -\frac{2/5 + 1/3\,q^{-1/3}(1+q^{1/3})^{-1}}{0.6 + q^{-2/3}\ln(1+q^{1/3})} \right]
\end{equation}
In the limiting case where $q \rightarrow 0$ (when the accretor is much heavier than the donor star; for example in a
soft {X}-ray transient system hosting a black hole):
\begin{equation}
   \lim _{q \rightarrow 0} \,\zeta _L = -5/3
\end{equation}
The behavior of $\zeta _L(q,\beta)$ for different {X}-ray binaries is plotted in Fig.~\ref{zetafig}.
This figure is quite useful to get an idea of the stability of the mass transfer when
comparing with  $\zeta$ of the donor star. We see that in general the Roche-lobe, $R_L$ increases
($\zeta _L <0$) when material is transfered from a relatively light donor to a heavy accretor ($q<0$).
In this situation the mass transfer will be stable.
Correspondingly $R_L$ decreases ($\zeta _L >0$) when material is transfered from a heavy donor to a
lighter accretor ($q>0$). In this case the mass transfer has the potential to be dynamically unstable.
This behavior can be understood from the bottom panel of Fig.~\ref{zetafig} where we
plot:
\begin{equation}
   -\partial\ln (a)/\partial\ln (q) = 2 + \frac{q}{q+1} + q\frac{3\beta -5}{q(1-\beta)+1}
\end{equation}
as a function of $q$. The sign of this quantity is important
since it tells whether the orbit expands or contracts in response to mass transfer (note $\partial q <0$).
It is noticed that the orbit always expands when $q<1$ and it always decreases when $q>1.28$
[solving $\partial\ln (a)/\partial\ln (q) = 0$ for $\beta=1$ yields $q=(1+\sqrt{17})/4 \approx 1.28$].
If $\beta > 0$ the orbit can still expand for $1 < q < 1.28$. There is a point at  $q=3/2$ where
$\partial\ln (a)/\partial\ln (q) = 2/5$, independent of $\beta$. It should also be mentioned for
the curious reader that if $\beta >0$ then, in some cases, it is actually possible for a binary
to decrease its separation, $a$ while increasing $P_{\rm orb}$ at the same time!

\subsection{Response of the mass-losing star -- the effect on the binary}
The radius of the mass-losing donor star will respond to mass loss.
Therefore, in order to obtain a full stability analysis of the mass-transfer process 
it is important to know whether or not the donor star expands (or contracts) in
response to mass loss. This is determined by the stellar structure (i.e. temperature gradient
and entropy) of the envelope at the onset of the RLO.

\subsubsection{Donor stars with a radiative envelope}
Donor stars with radiative (or only slightly convective) envelopes will usually shrink (or keep a 
roughly constant radius) in response to mass loss. Therefore, these stars will give rise to a dynamically
stable mass-transfer phase if the mass ratio, $q$ is not too large. Calculations by Tauris \& Savonije~(1999)
show that all LMXBs with donor stars $M_2 \le 1.8\,M_{\odot}$ and a neutron star will have a stable mass transfer phase.
Podsiadlowski, Rappaport \& Pfahl~(2002) find an upper limit for stability of $2.0\,M_{\odot}$. 
Stars with radiative envelopes are not very evolved and hence they are only found as donor stars in
systems with short orbital periods ($P_{\rm orb} \la 20$~days) at the onset of the RLO. 
Recently it was demonstrated in detail, that even IMXB systems with radiative donor stars $2 < M_2/M_{\odot} < 5$
are able to survive extreme mass transfer on a subthermal timescale (Tauris, van~den~Heuvel \& Savonije 2000).

\subsubsection{Donor stars with a convective envelope}
The thermal response of a donor star with a deep convective envelope is much more radical.
It expands rapidly in response to mass loss due to the super-adiabatic temperature gradient in its
giant envelope. This is clearly an unstable situation
if the Roche-lobe does not grow accordingly. For systems with $q\ga 1.5$ (heavy donors) 
the orbital shrinking is so efficient 
that, in combination with an expanding convective donor, it always leads
to the formation of a common envelope and (dynamically unstable) spiral-in evolution. 
Hence, this is most likely the destiny for all HMXBs.

\section{Common envelope (CE) evolution}
\label{CE}
A very important stage of the evolution of close binaries is the formation of a
common envelope. This phase is accompanied by the creation of a drag-force, arising from the motion of the
companion star through the envelope of the evolved star, which leads to
dissipation of orbital angular momentum (spiral-in process) and deposition of
orbital energy in the envelope. Hence, the global outcome of a CE-phase is
reduction of the binary separation and often ejection of the envelope.
There is strong evidence of orbital shrinkage (as brought about by frictional
torques in a CE-phase) in some observed close binary pulsars and white dwarf
binaries, e.g. PSR~1913+16, PSR~J1756--5322 and L~870--2.
In these systems it is clear that the precursor of the last-formed
degenerate star must have achieved a radius much larger than the current orbital
separation. 
There are many uncertainties involved in calculations of the spiral-in phase
during the CE evolution. The evolution is often tidally unstable and the angular
momentum transfer, dissipation of orbital energy and structural changes of the
donor star take place on very short timescales ($\sim\!10^3$ yr). A complete
study of the problem requires very detailed multi-dimensional
hydrodynamical calculations.
For a general review on common envelopes,
see e.g. Iben \& Livio~(1993) and Taam \& Sandquist~(2000).

A simple estimation of the reduction of the orbital separation can be found by simply
equating the binding energy of the envelope of the (sub)giant donor to the
required difference in orbital energy (before and after the CE-phase). Following the formalism
of Webbink~(1984) and de~Kool~(1990), 
let $0<\eta_\mathrm{CE}<1$ describe the efficiency of ejecting the envelope, i.e.
of converting orbital energy into the kinetic energy that provides the outward
motion of the envelope: $E_\mathrm{env} \equiv \eta_\mathrm{CE} \,\, \Delta
E_\mathrm{orb}$ or, 
\begin{eqnarray}
  \frac {G M_\mathrm{donor} M_\mathrm{env}}
    {\lambda\, a_\mathrm{i} \,r_\mathrm{L}} & \equiv & \eta_\mathrm{CE}
    \left[ \frac {G M_\mathrm{core} M_\mathrm{1}} {2\, a_\mathrm{f}} -
    \frac {G M_\mathrm{donor} M_\mathrm{1}} {2 \,a_\mathrm{i}} \right]
    \label{env_kool}
\end{eqnarray}
yielding the ratio of final (post-CE) to initial (pre-CE) orbital separation:
\begin{eqnarray}
  \frac {a_\mathrm{f}} {a_\mathrm{i}} & = &
  \frac {M_\mathrm{core} \,M_\mathrm{1}} {M_\mathrm{donor}}\;
  \frac {1} {M_\mathrm{1} + 2 M_\mathrm{env}/(\eta_\mathrm{CE}\,\lambda \,r_\mathrm{L})}
\end{eqnarray}
where $M_\mathrm{core}=M_\mathrm{donor}-M_\mathrm{env}$; $r_\mathrm{L} = R_\mathrm{L} / a_\mathrm{i}$
is the dimensionless Roche-lobe radius of the donor star so that 
$a_\mathrm{i}\,r_\mathrm{L} = R_\mathrm{L} \approx R_\mathrm{donor}$ and $\lambda$ is a parameter
which depends on the stellar mass-density distribution, and consequently also on the evolutionary
stage of the star. The orbital separation of the
surviving binaries is quite often reduced by a factor of $\sim\!100$ as a result of the spiral-in.
If there is not enough orbital energy available to eject the envelope the stellar components will merge
in this process.

\subsection{The binding energy of the envelope}
The total binding energy of the envelope to the core is given by:
\begin{eqnarray}
  E_\mathrm{bind} & = & - \int_{M_\mathrm{core}}^{M_\mathrm{donor}}
  \frac {G M(r)} {r} dm + \alpha_\mathrm{th} \int_{M_\mathrm{core}}^{M_\mathrm{donor}} U dm
  \label{env_han}
\end{eqnarray}
where the first term is the gravitational binding energy and $U$ is the internal
thermodynamic energy. The latter involves the basic thermal energy for a simple
perfect gas ($3 \Re T / 2 \mu$), the energy of radiation ($a T^{4} / 3 \rho$),
as well as terms due to ionization of atoms and dissociation of molecules
and the Fermi energy of a degenerate electron gas (Han~et~al. 1994,$\,$1995).
The value of $\alpha_\mathrm{th}$ depends on the details of the ejection
process, which is very uncertain. A value of $\alpha_\mathrm{th}$ equal to 0 or
1 corresponds to maximum and minimum envelope binding energy, respectively.
By simply equating Eqs.~(\ref{env_kool}) and (\ref{env_han})
one is able to calculate the parameter $\lambda$ for different evolutionary stages of a
given star. 

\begin{figure}
  \centering
    \centerline{\resizebox{6.5cm}{!}{\includegraphics{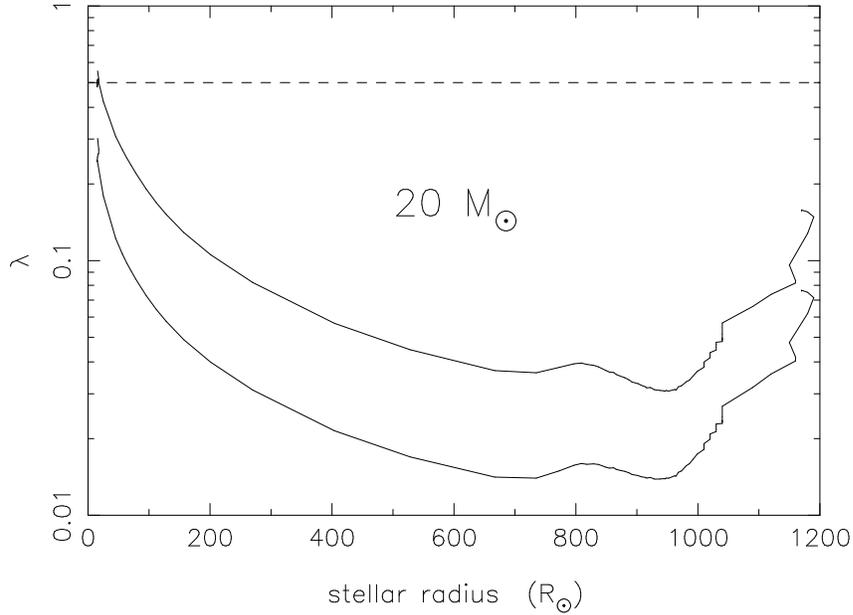}}}
  \caption{The $\lambda$-parameter for a $20\,M_{\odot}$ star as a function of stellar radius. The upper curve includes internal thermodynamic
               energy ($\alpha _{\rm th}=1$) whereas the lower curve is based on the sole gravitational binding energy ($\alpha _{\rm th}=0$)
               -- see Eq.~\ref{env_han}. There is a factor $\sim$2 in difference between the $\lambda$-curves in
               accordance with the virial theorem. It is a common misconception to use a constant value
               of $\lambda = 0.5$ (marked by the dashed line). See text for details.}
  \label{lambda20}
\end{figure}
Dewi \& Tauris~(2000) and Tauris \& Dewi~(2001) were the first to publish detailed calculations on
the binding energy of the envelope to determine the $\lambda$-parameter (however, see also Bisscheroux 1999).
Dewi \& Tauris (2000) investigated stars with masses $3-10\,M_{\odot}$ and found that
while $\lambda < 1$ on the RGB, $\lambda \gg 1$ on the AGB (especially for stars with $M<6\,M_{\odot}$).
Hence, the envelopes of these donor stars on the AGB are easily ejected; with only a relatively
modest decrease in orbital separation resulting from the spiral-in.
For more massive stars ($M>10\,M_{\odot}$) $\lambda < 0.1-0.01$ (see Fig.~\ref{lambda20}) and the internal energy is not very dominant 
(Dewi \& Tauris~2001; Podsiadlowski, Rappaport \& Han~2002).
This result has the important consequence that many HMXBs will
not produce double neutron star systems because they coalesce during their subsequent CE-phase
(leading to a relatively small merging~rate of double neutron star systems, as shown by Voss \& Tauris~2003).\\
It should be noticed that the exact determination of $\lambda$ depends on how the core boundary is defined
(see Tauris \& Dewi 2001 for a discussion). 
For example, if the core boundary (bifurcation point of envelope ejection in a CE)
of the $20\,M_{\odot}$ in Fig.\ref{lambda20} is moved out by $0.1\,M_{\odot}$ then $\lambda$ is typically increased
by a factor of $\sim\! 2$.

\subsubsection{The question of hypercritical accretion onto a NS in a CE}
It has been argued that a neutron star spiralling inwards in a common envelope might experience
hypercritical accretion and thereby collapse into a black~hole 
(e.g. Chevalier~1993; Brown~1995; Bethe \& Brown 1998; Fryer, Woosley \& Hartmann 1999). 
However, this idea seems difficult to conciliate with
the observations of a number of very tight-orbit binary pulsars. For example, the systems PSR J1756--5322 and PSR~B0655+64 
which have massive {CO} white dwarf companions ($\sim\!0.7\,M_{\odot}$ and $\sim\!1\,M_{\odot}$) with orbital periods
of only 0.45 and 1.06~days, respectively. From an evolutionary point of view there is no other way to produce such 
systems apart from a CE and spiral-in phase with donor masses between $3\,M_{\odot}$ and $6\,M_{\odot}$. 
From a theoretical point of view it has been argued that the hypercritical accretion can be
inhibited by rotation (Chevalier 1996) and strong outflows from the accretion disk (Armitage \& Livio 2000).
Finally, it should be noticed that the masses of neutron stars determined in all of the four detected Galactic double
neutron star systems are quite similar and close to a value of $1.4M_{\odot}$.
Hence, in all these cases it is clear that the first-born neutron star did not accrete any
significant amount of material from the common envelope.

\section{(Asymmetric) supernova explosions in close binaries}
After a close binary star has lost its {H}-envelope, and possibly also its {He}-envelope,
during RLO and/or CE evolution it will collapse and explode in a supernova (SN) if it is massive enough.
The critical threshold mass for a helium star to form a neutron star is about $2.8\,M_{\odot}$
(and somewhat lower for a CO-star, i.e. a helium star which has lost its helium envelope via case~BB RLO).
This value corresponds roughly to an initial mass of $10\,M_{\odot}$ ($\sim\!10\,M_{\odot}$ for case~C,
and $\sim\!12\,M_{\odot}$ for case~B/A RLO). If the core mass is below this critical threshold mass
the star contracts, possibly after a second phase of RLO, and settles peacefully as a cooling white dwarf.
If $M_{\rm He} \ga 8\,M_{\odot}$ the supernova leaves behind a black hole.
All observed neutron stars in binary pulsars seem to have
been born with a canonical mass of $1.3-1.4\,M_{\odot}$. Neutron stars may afterwards in LMXBs possibly accrete up to
$\sim\!1\,M_{\odot}$ before collapsing further into a black hole (see Sect.~\ref{NSmass}). 
In very massive {X}-ray binaries, neutron stars probably can be born with higher masses $\sim\!1.8\,M_{\odot}$
(Barziv et~al. 2001).

As mentioned in the introduction, there is firm evidence that most newborn neutron stars receive a momentum
kick at birth which gives rise to the high velocities observed (typically $\sim\!400$~km$\,$s$^{-1}$),
although there also may well be a smaller fraction $\sim \!10-20\,$\% that receive kicks $\la 50$~km~s$^{-1}$
in order to explain the observed neutron stars in globular clusters and the population of very wide-orbit
{X}-ray binaries (Pfahl et~al. 2002). 
It is still an open question whether or not black holes also receive a kick at birth.
At least in some cases there are observational indications of mass ejection during their formation, 
as in any successful supernova explosion (Iwamoto~et~al.~1998), and the recent
determination of the run-away velocity ($112\pm18$~km~s$^{-1}$) of the black hole binary GRO~J1655--40
(Mirabel et~al.~2002) seems to suggest that also the formation of black holes
are accompanied by a kick.

In an excellent paper Hills~(1983) calculated the dynamical
consequences of binaries surviving an asymmetric SN. Tauris \& Takens~(1998) generalized the problem and derived
analytical formulas for the velocities of stellar components ejected from disrupted binaries and also included
the effect of shell impact on the companion star. 
If the collapsing core of a helium star results in a supernova explosion it is a good approximation
that the collapse is instantaneous compared with $P_{\rm orb}$. Here we summarize a few important equations.
The orbital energy of a binary is given by:
\begin{equation}
   E_{\rm orb} = -\frac{G M_1 M_2}{2\,a} = -\frac{G M_1 M_2}{r} + \textstyle\frac{1}{2}\mu v_{\rm rel}^2
\end{equation}
where $r$ is the separation between the stars at the moment of explosion; $\mu$ is the reduced mass
of the system and $v_{\rm rel}=\sqrt{GM/r}$ is the relative velocity of the two stars in a circular
pre-SN binary. The change of the semi-major axis as a result of the SN is then given by
(Flannery \& van~den~Heuvel 1975):
\begin{equation}
\label{SNeq}
  \frac{a}{a_0} = \left[ \frac{1-(\Delta M/M)}{1-2(\Delta M/M) - (w/v_{\rm rel})^2 -2\cos\theta \,(w/v_{\rm rel})} \right]
\end{equation} 
where $a_0=r$ and $a$ are the initial and final semi-major axis, respectively; $\Delta M$ is the amount of matter lost
in the SN; $w$ is the magnitude of the kick velocity and $\theta$ is the direction of the kick relative to the orientation
of the pre-SN velocity. The orientation of the kick magnitude is probably completely uncorrelated with respect to the
orientation of the binary -- the escaping neutrinos from deep inside the collapsing core are not aware that they are
members of a binary system (see, however, Pfahl~et~al.~(2002) for the hypothesis that the kick {\em magnitude} may depend
on the pre-SN history of the collapsing core). For each binary there exists a critical angle, $\theta _{\rm crit}$ for which a
SN with $\theta  < \theta _{\rm crit}$ will result in disruption of the orbit (i.e. if the denominator of Eq.~\ref{SNeq} is less than zero).

\noindent The sudden mass loss in the SN affects the bound orbit with an eccentricity:
\begin{equation}
  e=\sqrt{1+\displaystyle\frac{2E_{\rm orb}L_{\rm orb}^2}{\mu G^2 M_1^2 M_2^2}}
\end{equation} 
where the orbital angular momentum can be derived from (see also Eq.~\ref{Jorb}):
\begin{equation}
  L_{\rm orb} = |\vec{r}\times \vec{p}\,| = r\,\mu\sqrt{(v_{\rm rel} + w\cos\theta )^2 + (w\sin\theta\,\sin\phi )^2}
\end{equation} 
Note, in the two equations above $v_{\rm rel}$ is the pre-SN relative velocity of the two stars, whereas
the stellar masses and $\mu$ now refer to the post-SN values. 

Systems surviving the SN will receive a recoil velocity from the combined effect of instant 
mass loss and a kick. One can easily find this velocity, $v_{\rm sys}$ from conservation of momentum. Let us consider 
a star with mass $M_{\rm core}$ collapsing to form a neutron star with mass $M_{\rm NS}$ and hence:
\begin{equation}
   v_{\rm sys} = \sqrt{\Delta P_x^2 +\Delta P_y^2 + \Delta P_z^2}/(M_{\rm NS}+M_2)
\end{equation} 
where the change in momentum is:
\begin{eqnarray}
   \Delta P_x & = & M_{\rm NS}(v_{\rm core}+w\cos\theta) - M_{\rm core}v_{\rm core}\nonumber\\
   \Delta P_y & = & M_{\rm NS}w\sin\theta\cos\phi\\
   \Delta P_z & = & M_{\rm NS}w\sin\theta\sin\phi\nonumber
\end{eqnarray}
and where $M_2$ is the unchanged mass of the companion star; $v_{\rm core}$ is the pre-SN velocity of the collapsing 
core, in a center of mass reference frame, and $\phi$ is the angle between the projection of $\vec{w}$ onto
a plane $\perp$ to $\vec{v}_{\rm core}$ (i.e. $w_y = w\sin\theta\cos\phi$). Beware, if the post-SN periastron distance,
$a(1-e)$ is smaller than the radius of the companion star then the binary will merge.\\
It is important to realize the difference
between {\em gravitational} mass (as measured by an observer) and {\em baryonic} mass of a neutron star. 
The latter is $\sim\!15\,$\% larger for a typical equation-of-state. When considering dynamical effects
on binaries surviving a SN this fact is often (almost always) ignored!  

\section{Evolution of LMXBs -- formation of millisecond pulsars}
\begin{figure}
  \centering
 \centerline{\resizebox{23cm}{!}{\includegraphics{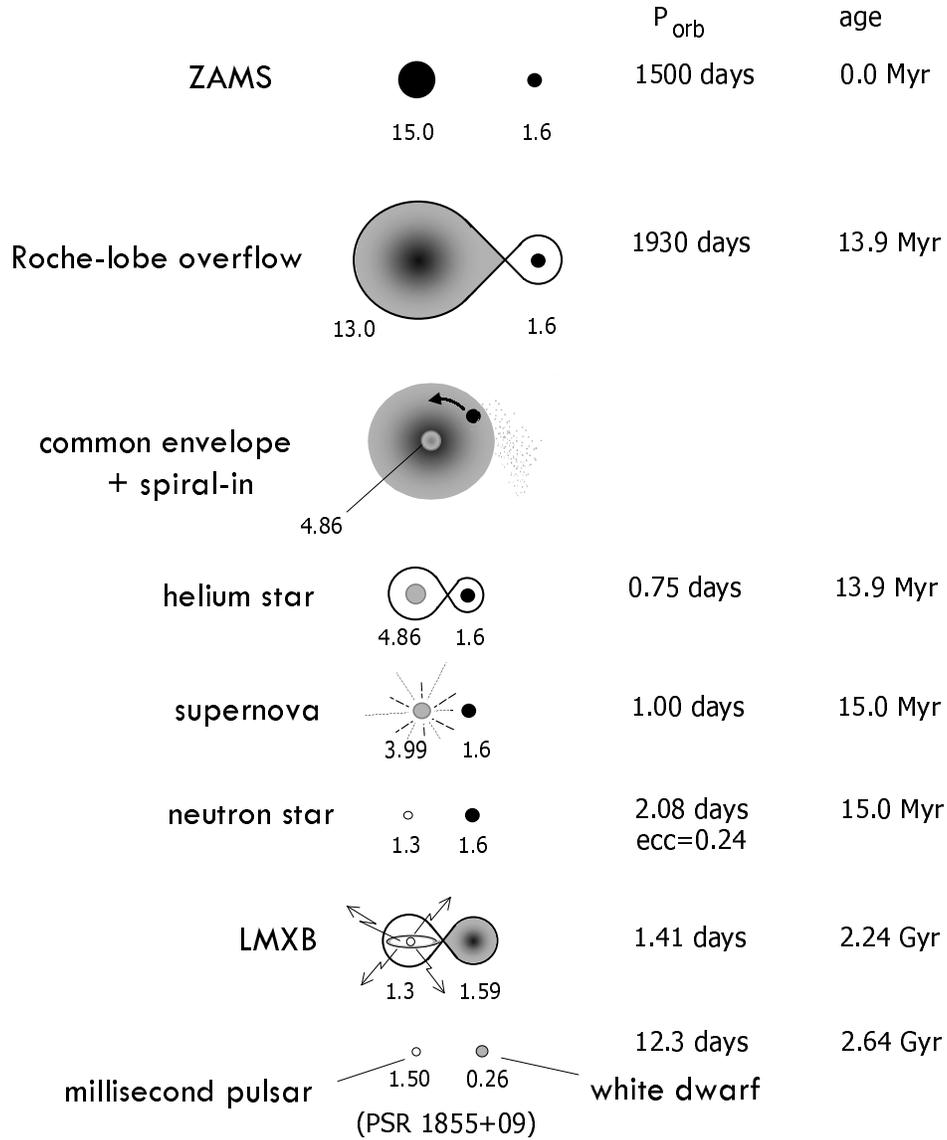}}}
  \caption{Cartoon depicting the evolution of a binary system eventually leading to an LMXB and finally the formation 
           of a binary millisecond pulsar. Parameters governing the specific orbital angular momentum of ejected matter,
           the common envelope and spiral-in phase, the asymmetric
           supernova explosion and the stellar evolution of the naked helium star all have a large impact on the exact evolution.
           Parameters are given for a scenario leading to the formation of the observed binary millisecond pulsar
           PSR~1855+09. The stellar masses given are in solar units.
           } 
  \label{LMXB-cartoon}
\end{figure}
\begin{figure}[t]
  \centering
 \centerline{\resizebox{12.5cm}{!}{\includegraphics{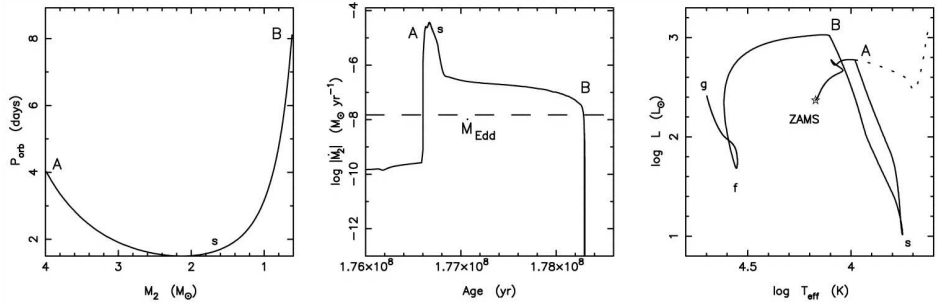}}}
  \caption{The evolution of an IMXB leading to the formation of a BMSP with a {CO} WD companion
           in a close orbit. The initial configuration was a $4\,M_{\odot}$ donor star and a neutron star
           with an orbital period of 4~days. The mass-transfer phase is between points~A and B.
           Between points~f and g helium is burning in the core of the stripped companion star.
           After Tauris, van~den~Heuvel \& Savonije~(2000).}
  \label{IMXB}
\end{figure}

Fig.~\ref{LMXB-cartoon} depicts the formation of an LMXB and millisecond pulsar system. 
There are now more than 40 binary millisecond pulsars (BMSPs) known in the Galactic disk. They can be
roughly divided into three observational classes (Tauris~1996): class~A contains the wide-orbit ($P_{\rm orb} > 20$~days)
BMSPs with low-mass helium white dwarf companions ($M_{\rm WD} < 0.45\,M_{\odot}$), whereas the close-orbit
BMSPs ($P_{\rm orb} \la 15$~days) consist of systems with either low-mass helium white dwarf companions
(class~B) or systems with relatively heavy {CO}/{O}-{Ne}-{Mg} white dwarf companions (class~C).
The latter class evolved through a phase with significant loss of angular momentum (either common
envelope evolution or extreme mass transfer on a subthermal timescale) 
and descends from IMXBs with donors: $2 < M_2/M_{\odot} < 8$, see Fig.~\ref{IMXB}.
The single MSPs are believed to originate from tight class~B systems where the companion has been destroyed
or evaporated -- either from {X}-ray irradiation when the neutron star was accreting, or in the form of
a pulsar radiation/wind of relativistic particles (e.g. van~den~Heuvel \& van~Paradijs~1988; 
Ruderman, Shaham \& Tavani~1989; Podsiadlowski~1991; Ergma \& Fedorova~1991; Shaham~1992; Tavani~1992).
Observational evidence for this scenario is found in eclipsing MSPs with ultra light companions --
e.g. PSR 1957+20 ($P_{\rm orb}=0.38$~day; $M_2\simeq 0.02\,M_{\odot}$) and the planetary pulsar:
PSR 1257+12 (Wolszczan~1994).

For LMXBs
it has been shown by Pylyser \& Savonije~(1988,$\,$1989) that an orbital bifurcation period separates the formation
of converging systems (which evolve with decreasing $P_{\rm orb}$ until the mass-losing component becomes degenerate
and an ultra-compact binary is formed) from the diverging systems (which finally evolve with increasing $P_{\rm orb}$
until the mass-losing star has lost its envelope and a wide detached binary is formed).
This important bifurcation period is about 2--3 days depending on the strength of the magnetic braking torque.

\subsection{Formation of wide-orbit binary millisecond pulsars}
In LMXBs with initial $P_{\rm orb} >2$ days the mass transfer is driven by internal thermonuclear
evolution of the donor star since it evolves into a (sub)giant before loss of orbital angular momentum
dominates. These systems have been studied by Webbink, Rappaport \& Savonije~(1983), Taam~(1983), Savonije~(1987), 
Joss, Rappaport \& Lewis~(1987), Rappaport~et~al.~(1995) and recently
Ergma, Sarna \& Antipova~(1998), Tauris \& Savonije~(1999) and Podsiadlowski, Rappaport \& Pfahl~(2002).
For a donor star on the red giant branch (RGB) the growth in core mass is directly related to the luminosity,
as this luminosity is entirely generated by hydrogen shell burning. As such a star, with a small compact core surrounded
by an extended convective envelope, is forced to move up the Hayashi track its luminosity increases strongly
with only a fairly modest decrease in temperature. Hence one also finds a relationship
between the giant's radius and the mass of its degenerate helium core -- almost entirely independent
of the mass present in the hydrogen-rich envelope (see Table~\ref{coremass}).
\begin{table}[b]
  \caption{Stellar parameters for giant stars with $R=50.0\,R_{\odot}$} 
    \begin{tabular}{l|llll}
     \hline \hline
     $M_{\rm initial}/M_{\odot}$ & $1.0^{*}$ & $1.6^{*}$ & $1.0^{**}$ & $1.6^{**}$\\ 
     \hline
     log $L/L_{\odot}$ & 2.644 & 2.723 & 2.566 & 2.624\\
     log $T_{\rm eff}$ & 3.573 & 3.593 & 3.554 & 3.569\\
     $M_{\rm core}/M_{\odot}$ & 0.342 & 0.354 & 0.336 & 0.345\\
     $M_{\rm env}/M_{\odot}$ & 0.615 & 1.217 & 0.215 & 0.514\\
     \hline \hline
    \end{tabular}
    \\
    $^{*}$ Single star ({X}=0.70, {Z}=0.02 and $\alpha = 2.0, \;\delta _{\rm ov}=0.10$).\\
    $^{**}$ Binary star (at onset of RLO: $P_{\rm orb} \simeq 60$ days and $M_{\rm NS}=1.3\,M_{\odot}$).\\
    After Tauris \& Savonije~(1999).\\
  \label{coremass}
\end{table}
\begin{figure}
  \centering
 \centerline{\resizebox{7cm}{!}{\includegraphics{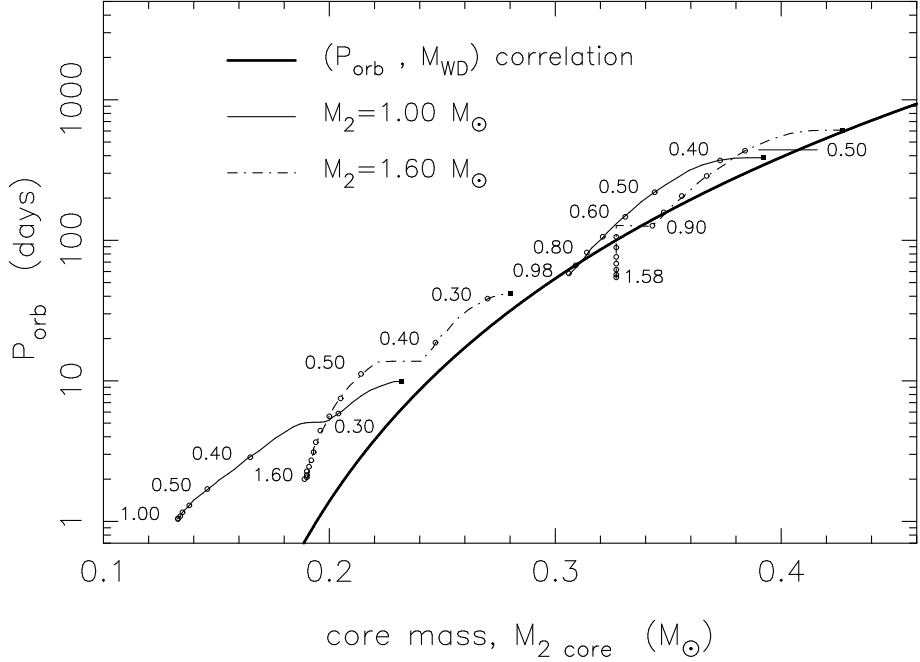}}}
  \caption{Evolutionary tracks of four LMXBs showing $P_{\rm orb}$ as a function of $M_{\rm core}$ of the
   donor star. The initial donor masses were 1.0 and $1.6\,M_{\odot}$ (each calculated at two different initial $P_{\rm orb}$)
   and the initial neutron star mass was $1.3\,M_{\odot}$. 
   The total mass of the donors during the evolution is written along the tracks.
   At the termination of the mass-transfer process the donor only has a tiny ($\le 0.01\,M_{\odot}$) 
   {H}-envelope and the end-points of the evolutionary tracks are located near the curve representing
   the ($P_{\rm orb}, M_{\rm WD}$) correlation for BMSPs. After Tauris \& Savonije~(1999).}
  \label{PM-rel}
\end{figure}
\hspace{-0.2cm}It has also been argued that the core-mass determines the rate of
mass-transfer (Webbink, Rappaport \& Savonije 1983). In the scenario under consideration,
the extended envelope of the giant is expected to fill its Roche-lobe until termination of the mass transfer.
Since the Roche-lobe radius, $R_L$ only depends on the masses and separation between the two stars it is clear
that the core mass, from the moment the star begins RLO, is uniquely correlated with $P_{\rm orb}$ of the system.
Thus also the final orbital period ($\sim\!2$ to $10^3$~days) is expected to be a function of the mass of the resulting
white dwarf companion (Savonije~1987). Tauris \& Savonije~(1999) calculated the expected ($P_{\rm orb}, M_{\rm WD}$) 
correlation in detail and found an overall best fit:
\begin{equation}
  M_{\rm WD} = \left ( \frac{P_{\rm orb}}{b}\right) ^{\textstyle 1/a} + \,c
\end{equation}
where, depending on the chemical composition of the donor,
\begin{equation}
  (a,b,c) = \, \left\{ \begin{array}{llll}
   4.50 & 1.2\times 10^5 & 0.120 & \mbox{\hspace{0.1cm}Pop.I}\\ 
   4.75 & 1.1\times 10^5 & 0.115 & \mbox{\hspace{0.1cm}Pop.I+II}\\ 
   5.00 & 1.0\times 10^5 & 0.110 & \mbox{\hspace{0.1cm}Pop.II}\\ 
               \end{array}
         \right.
\end{equation}
Here $M_{\rm WD}$ is in solar mass units and $P_{\rm orb}$ is measured in days.
The fit is valid for BMSPs with helium white dwarfs companions and $0.18 \le M_{\rm WD}/M_{\odot} \le 0.45$.
The formula depends slightly on the adopted value of the convective mixing-length parameter.
It should be noted that the correlation is {\em independent} of $\beta$ (the fraction of the transferred material
lost from the system -- see Sect.~\ref{OAMB}), the mode of the mass loss and the strength of the magnetic braking torque, since
the relation between giant radius and core mass of the donor star remains unaffected by the exterior
stellar conditions governing the process of mass transfer.
However, for the {\em individual} binary $P_{\rm orb}$ and $M_{\rm WD}$ do depend on these parameters. 
In Fig.~\ref{PM-rel} we have plotted a theoretical ($P_{\rm orb}, M_{\rm WD}$) correlation and 
also plotted evolutionary tracks calculated for four LMXBs.
Although clearly the class of BMSPs with helium white dwarf companions is present, 
the estimated masses of the BMSP white dwarfs are quite
uncertain, since they depend on the unknown orbital inclination angle and the pulsar mass, and
no clear observed ($P_{\rm orb}, M_{\rm WD}$) correlation has yet been established from the current observations. 
In particular there may be a discrepancy for the BMSPs with $P_{\rm orb} \ga 100$~days 
(Tauris~1996). 

\subsection{Formation of close-orbit binary millisecond pulsars}
In LMXBs with initial $P_{\rm orb} <2$ days the mass transfer is driven by loss of angular momentum due
to magnetic braking and gravitational wave radiation.
The evolution of such systems is very similar to the evolution of CVs -- see e.g. Spruit \& Ritter~(1983);
Verbunt \& van~den~Heuvel~(1995); Ergma, Sarna \& Antipova~(1998) and Warner \& Kuulkers (this book). 

\subsection{Masses of binary neutron stars}
\label{NSmass}
In general, the masses of binary pulsars can only be estimated from their observed mass function, which
depends on the unknown orbital inclination angle. Only in a few tight systems is it possible to directly
measure post-Newtonian parameters (e.g. the general relativistic Shapiro delay) which yield precise
values of the stellar masses (see Taylor \& Weisberg 1989). 
For example, in the double neutron star system PSR~1913+16 the (gravitational) masses
are known to be 1.441 and $1.387\,M_{\odot}$. Although the majority of the (rough) estimated pulsar masses
may be consistent with the canonical value of $\sim\!1.4\,M_{\odot}$ (Thorsett \& Chakrabarty 1999),
one could still expect a spread in neutron star masses from an evolutionary point of view.
The recycled pulsars in double neutron star systems, for example, did not have a chance to accrete
much material because of the short-lived common envelope and spiral-in phase that these systems evolved through,
according to the standard model.
Assuming all neutron stars to be born with a mass of $1.3\,M_{\odot}$, Tauris \& Savonije~(1999) demonstrated
that an ($P_{\rm orb}, M_{\rm NS}$) anti-correlation would be expected for millisecond pulsars as a simple consequence of the
interplay between mass-transfer rate (and thus accretion rate), orbital period and the
evolutionary status of the LMXB donor star at the onset of the RLO. However, since this model
predicted rather massive ($>2\,M_{\odot}$) neutron stars in binary millisecond pulsar systems with
$P_{\rm orb} \la 30$~days, it failed to explain the mass of PSR~B1855+09 ($P_{\rm orb}=12.3$~days)
which is known to be $< 1.55\,M_{\odot}$
from constraints on its Shapiro delay. The authors concluded that this was a proof for the fact that a
large amount of matter {\it must} be lost from the LMXB even for sub-Eddington accretion -- probably as a result
of either accretion disk instabilities (Pringle~1981; van~Paradijs~1996) or the so-called propeller effect
(Illarionov \& Sunyaev 1985). 

The maximum mass of a neutron star depends on the equation-of-state for dense matter.
Barziv~et~al.~(2001) reported that the HMXB Vela~X-1 has a neutron star mass
$\sim\!1.86\,M_{\odot}$. Furthermore, some kHz~QPO sources
are claimed to host heavy ($>2\,M_{\odot}$) neutron stars. If this is the case, then all soft EOS, 
including some kaon condensation EOS (Brown \& Bethe 1994; Bethe \& Brown 1995), can be ruled out.

\section{Evolution of HMXBs}
\label{HMXB_evol}
\subsection{Formation of double neutron star/blcak hole binaries}
The formation of a HMXB requires two relatively massive stars ($>12\,M_{\odot}$). Alternatively the secondary ZAMS
star can be less massive initially, as long as it gains enough material from the primary star, so that it will
later end up above the threshold mass for undergoing a supernova explosion (like the primary star). 
The first mass transfer phase, from the primary to the secondary star, is usually assumed to be dynamically stable
(semi-conservative) if the
mass ratio at the onset of the RLO is not too extreme. However, later on {\em all} HMXBs end up in a common envelope phase,
as the neutron star (or low-mass black hole) is engulfed by the extended envelope of its companion, in an orbit
which is rapidly shrinking due to heavy loss of orbital angular momentum.
As discussed earlier in Sect.~\ref{heliumevol},
stellar winds of massive stars, as well as naked helium cores (Wolf-Rayet stars), are some of the most uncertain
aspects of the modelling of HMXB evolution. The physical conditions which determine the formation of
a neutron star versus a black hole are also still quite unknown. It may well be that core mass is
not the only important factor to determine the outcome. Magnetic field and spin of the collapsing core
could also play a major role (Ergma \& van~den~Heuvel 1998). Furthermore, it seems clear from observations
that there is an overlap in the mass range for making a neutron star versus a black hole.\\

\subsection{Gravitational waves and merging NS/BH binaries} 
\begin{figure}
  \centering
 \centerline{\resizebox{6cm}{!}{\includegraphics{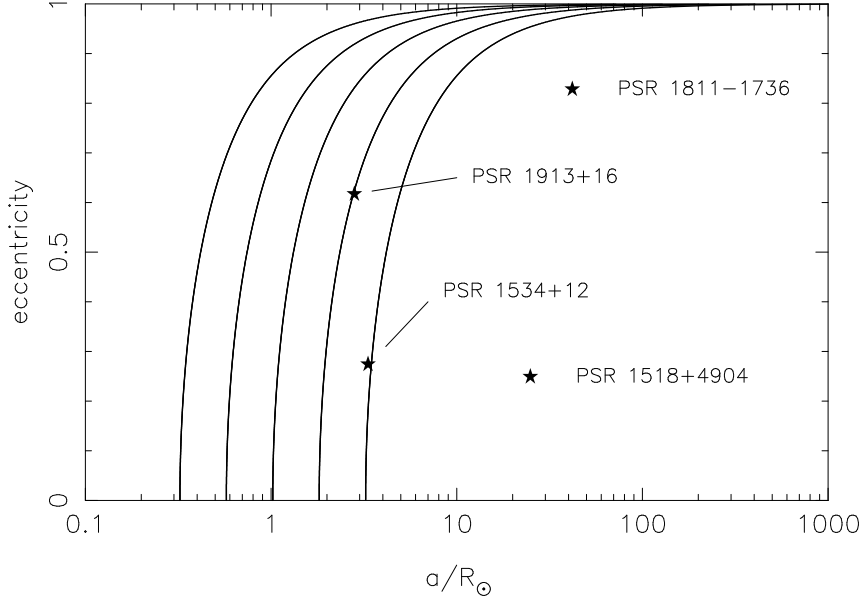}}}
  \caption{Isochrones for the merging time of double neutron star binaries, as calculated by the authors.
           The curves correspond to values of (from left to right): $3\times 10^5$~yr, 3~Myr, 30~Myr, 300~Myr and
           3~Gyr, respectively. The four detected Galactic double neutron star systems are indicated with $\star$.}
  \label{mergingtime}
\end{figure}

The rate of energy loss as a result of gravitational wave radiation (GWR) is given by
(in the quadrupole approximation, $a \ll \lambda_{\rm gwr}$):
\begin{equation}
  L_{\rm gwr} = \left| \frac{dE}{dt}\right| = \frac{G}{5c^5} 
                \langle \,^{^{^{...}}}\!\!\!\!\!\!Q_{jk} \,^{^{^{...}}}\!\!\!\!\!\!{Q}_{jk} \rangle
                \; g(n,e) \; \simeq \frac{32 G^4}{5c^5}\frac{M^3\mu ^2}{a^5}\; f(e)
\end{equation}
where $Q$ denotes the quadrupole moment of the mass distribution; $M$ is the total mass of the system;
$\mu$ is the reduced mass and $f(e)$ is a function of the orbital eccentricity
(here we have disregarded the dependence on the harmonic number of the wave signal).
The energy loss due to GWR can only be subtracted from the orbital energy of the binary
and hence the orbital separation will decrease as:
\begin{equation}
  \dot{a}= \frac{GM\mu}{2E_{\rm orb}^2}\,\dot{E}_{\rm orb} \quad
  \left( a= -\frac{GM\mu}{2E_{\rm orb}} \;\; \wedge \;\; \dot{E}_{\rm orb}=-L_{\rm gwr} \right)
\end{equation}
For an eccentric binary: 
\begin{equation}
  \frac{1}{a}\frac{da}{dt} = -\frac{1}{E}\left.\frac{dE}{dt}\right|_{e=0}\, f(e)
  \quad \wedge \quad f(e) \simeq \frac{1+\frac{73}{24}e^2 + \frac{37}{96}e^4}{(1-e^2)^{7/2}}
\end{equation}
where the approximate fit for $f(e)$ above is given by Peters~(1964).\\
Now we have an expression for the rate of change in the orbital separation:
\begin{equation}
  \dot{a} \simeq -\frac{64 G^3}{5c^5}\frac{M^2\mu}{a^3}\,\frac{1+\frac{73}{24}e^2 + \frac{37}{96}e^4}{(1-e^2)^{7/2}}
\end{equation}
which can be transformed into an expression for the merging time, $\tau_{\rm gwr}$ (Peters 1964)
as a function of the initial values ($a_0, e_0$):
\begin{equation}
\tau_{\rm gwr} (a_0,e_0)=\frac{12}{19}\frac{C_0^{4}}{\beta}\times \int_{0}^{e_0}\frac{e^{29/19}[1+(121/304)e^{2}]^
{1181/2299}}{(1-e^{2})^{3/2}}de
\end{equation}
where 
\begin{equation}
C_0=\frac{a_0(1-e_0^{2})}{e_0^{12/19}}\,[1+(121/304)e_0^{2}]^{-870/2299} \;\; \wedge \;\; \beta =\frac{64G^3}{5c^5}M^2\mu
\end{equation}
This equation cannot be solved analytically and must be evaluated
numerically. 
The timescale is very dependent on both $a$ and $e$. Tight and/or eccentric orbits
spiral-in much faster than wider and more circular orbits -- see Fig.~\ref{mergingtime}.
For example, we find that the double neutron star system PSR~1913+16 
($P_{\rm orb}=7.75$~hr, $M_{\rm NS}=1.441$ and $1.387\,M_{\odot}$) 
with an eccentricity of 0.617 will merge in 302 Myr; if its orbit
was circular the merger time would be five times longer: 1.65~Gyr! 
For circular orbits the merging timescale can easily be found analytically: $\tau _{\rm gwr}^{\rm circ} = a_0^4/4\beta$.

\subsubsection{LISA/LIGO observations of signals from tight NS/BH binaries}
The double neutron star system PSR~B1913+16 has an orbital period of 7.75~hr and a distance to
the Earth of $\sim\!7$~kpc. Thus the signal from this source is too weak, and the wave frequency
too small, to be detected by LISA. However, it is clear from population synthesis studies 
that many NS-NS~binaries must exist with orbital periods of less than a few hours. Such a
system, relatively nearby, would be detected by LISA in the mHz range from its continuous emission of gravitational waves.\\
The amplitude (or strain) of gravitational waves emitted continuously from a tight binary is given by
the sum of the two polarizations of the signal:
\begin{eqnarray}
   h & = & \sqrt{\textstyle\frac{1}{2}[h^2_{+, max}+h^2_{\times, max}]}\;
          =\; \sqrt{\displaystyle\frac{16\pi\,G}{c^3 \omega_{\rm gwr}}\frac{L_{\rm gwr}}{4\pi\,d^2}} \nonumber \\ 
     & \simeq & 1\times 10^{-21}\;\left( \frac{M_{\rm chirp}}{M_{\odot}}\right) ^{5/3}
         \left( \frac{P_{\rm orb}}{1\, \rm{hr}} \right) ^{-2/3} \left( \frac{d}{1\, \rm{kpc}} \right) ^{-1}
\end{eqnarray}
Here the waves are assumed to be sinusoidal with angular frequency, $\omega_{\rm gwr}$ which is
$\sim\!2$ times the orbital angular frequency of the binary ($\Omega = 2\pi/P_{\rm orb}$), and
$M_{\rm chirp}\equiv \mu^{3/5}M^{2/5}$ is the so-called chirp mass of the system.

As a compact binary continues its inspiral, the gravitational waves will sweep upward in frequency
from about 10~Hz to $10^3$~Hz, at which point the compact stars will collide and coalesce. It is
this last 15~minutes of inspiral, with $\sim\!16\,000$ cycles of waveform oscillation, and the final
coalescence, that LIGO/VIRGO seeks to monitor. LIGO~I and LIGO~II are expected to detect
NS-NS inspiral events out to a distance of $\sim\!20$~Mpc and $\sim\!300$~Mpc, respectively, according
to recent estimates (Thorne~2001).
This corresponds to wave amplitudes of roughly $10^{-20} > h > 10^{-22}$. As a result of the much larger 
chirp~mass for the BH-BH mergers, such binaries will be detected out to a distance luminosity, 
$d_L \propto M_{\rm chirp}^{5/6}$ (Finn 1998) which is about
4 times larger. Hence, the ratio of detected event rates for BH-BH mergers relative to NS-NS mergers is
$4^3 \sim 64$ times larger than the corresponding ratio of such mergers in our Galaxy. As a result BH-BH mergers
are expected to be dominant for LIGO detectors as noted by Sipior \& Sigurdsson~(2002).

The cosmological implications of gravitational
wave observations of binary inspiral are also interesting to note (Schutz 1986; Finn 1997). 
Finally, it should be mentioned that LIGO~II is expected to detect burst signals from extra-galactic 
supernova explosions as well (Thorne~2001). Investigations of these signals may help to reveal the unknown progenitors of both short 
and long-duration gamma-ray bursts (GRBs). 

\subsubsection{Galactic merger rates of NS/BH binaries}
It is very important to constrain the local merging rate of NS/BH-binaries in order to predict
the number of events detected by LIGO. This rate can be determined either from binary population
synthesis calculations, or from observations of Galactic NS-NS systems (binary pulsars).
Both methods involve a large number of uncertainties (e.g. Kalogera~et~al.~2001 and references therein).
The current estimates for the Galactic merger rate of NS-NS systems are in the range $10^{-6} - 10^{-4}$~yr$^{-1}$.

In order to extrapolate the Galactic coalescence rate out to the volume of the universe accessible
to LIGO, one can either use a method based on star formation rates or a scaling based on the
B-band luminosities of galaxies. Using the latter method Kalogera~et~al.~(2001) found a scaling
factor of $(1.0-1.5)\times10^{-2}$~Mpc$^{-3}$, or equivalently, 
$\sim\!400$ for LIGO~I (out to 20~Mpc for NS-NS mergers). Since LIGO~II is expected to look out to a distance
of 300~Mpc (for NS-NS mergers), the volume covered by LIGO~II is larger by a factor of (300/20)$^3$ and thus the
scaling factor in this case, relative to the coalescence rates in the Milky Way, is about $1.3\times 10^6$.
Therefore, the expected rate of detections from NS-NS inspiral events is roughly between 2--200~yr$^{-1}$
for LIGO~II. However, LIGO~I will probably not detect any NS/BH inspiral event.  

Recent studies (Voss \& Tauris~2003 and Dewi, Pols \& van~den~Heuvel 2003) have included `real $\lambda$-values' 
for the CE-phase and this
results in a relatively low Galactic NS-NS merger rate (since fewer systems survive the CE evolution). 
However, Voss \& Tauris~(2003) also investigated BH-BH systems and find a relatively high Galactic merger rate of such systems, 
compared to NS-NS systems and mixed NS/BH systems, and estimate a BH-BH merging detection rate of 840~yr$^{-1}$ for LIGO~II.\\
One should be aware that compact mergers in globular clusters probably also contribute significantly
to the total merger rates (Portegies~Zwart \& McMillan 2000).

\section{Spin and $\vec{B}$-field evolution of accreting neutron stars}
\begin{figure}[t]
  \centering
    \centerline{\resizebox{12cm}{!}{\includegraphics{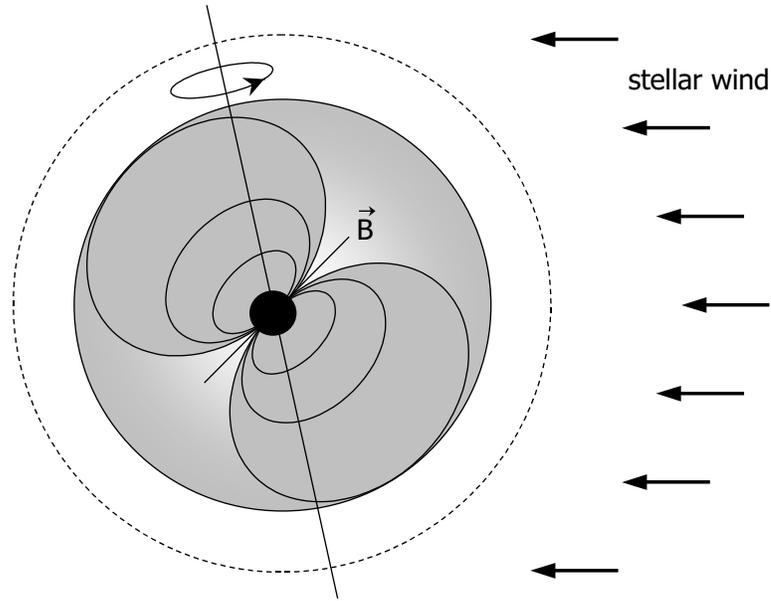}}}
  \caption{Illustration of the magnetosphere surrounding a wind-accreting neutron star. The rotation period,
           the magnetic field strength and the ram pressure of the wind determines whether or not
           accretion onto the neutron star surface is possible.
           The spin axis and the magnetic field axis are misaligned (thus: a pulsar). {X}-rays are
           emitted along the magnetic field axis as the pulsar accretes near its magnetic poles.}
  \label{magnetosphere}
\end{figure}
Most {X}-ray pulsars have spin periods between $10-1000$~sec. In persistent sources, spin periods of the
order of seconds are found only in systems where we have clear evidence
for the presence of an accretion disk from UV and optical observations. 
These are systems where (a large part of) the mass transfer
is due to RLO: the LMXBs and those few HMXBs in which the supergiant donors are just beginning to overflow
their Roche-lobes (Cen~X-3, SMC~X-1 and LMC~X-4). The latter ones are expected to be powered by a combination
of stellar wind and beginning atmospheric RLO (Savonije~1978,$\,$1983).
In most of these systems the {X}-ray pulsars show a secular decrease of their pulse period (spin-up)
on a relatively short timescale ($10^3 - 10^5$~yr). 
Although on short timescales episodes of spin-up and spin-down alternate,
the average trend is clearly that of spin-up (see e.g. Bildsten~et~al. 1997).
In sources which are purely wind-fed, the HMXBs with blue supergiant companions
that do not yet fill their Roche-lobe, the pulse periods are very long, and they vary erratically in time
showing no clear secular trends. This can be explained by the fact that the amount of angular momentum
carried by the supersonic winds is negligible, and eddies form in the wind downstream of the neutron star,
which alternately may feed co- and counter-rotating angular momentum to it (Taam \& Fryxell 1988;
Fryxell \& Taam 1988; Matsuda~et~al. 1992).

The accretion and spin-evolution of a neutron star in a binary system depends on a number of parameters:
the magnetodipole radiation pressure, the ram pressure of the companion star wind, 
the radius of gravitational capture, the location of the light cylinder,
the Alfv\'en radius, the co-rotation radius, the propeller effect and whether or not an accretion disk is formed
-- see e.g. Ghosh \& Lamb~(1979) and King elsewhere in this book for a detailed description. 
The old binary neutron stars, which have been spun-up and recycled in LMXBs,
reappear as observable millisecond {X}-ray and radio pulsars. Their so-called equilibrium
spin period is given by (see the spin-up line in the ($P,\dot{P}$)-diagram in Fig.~\ref{PPdot}): 
\begin{equation}
   P_{\rm eq} \propto \dot{M}_{\rm NS}^{-3/7}\,B^{6/7}\,R^{18/7}_{\rm NS}\,M^{-5/7}_{\rm NS} 
   \label{Peq}
\end{equation}
As long as $P>P_{\rm eq}$ accretion onto the neutron star is possible. 
Note, that while ordinary single pulsars have a typical lifetime of $10-100$~Myr, the recycled millisecond pulsars
have lifetimes of several Gyr and thus continue to lit the sky with beamed pulses for a Hubble-time. 
The reason for this is simply their low values of $\dot{P}$, i.e. relatively weak braking (or radiation-reaction) torques.

\subsection{The Eddington accretion limit}
The typical {X}-ray luminosities of the LMXBs and HMXBs are in the range $10^{35}-10^{38}$~erg~s$^{-1}$,
corresponding to mass accretion rates onto a neutron star in the range $10^{-11}-10^{-8}\,M_{\odot}$~yr$^{-1}$.
When the mass transfer rate exceeds $\dot{M}_{\rm Edd} \simeq 1.5\times 10^{-8}\,M_{\odot}$~yr$^{-1}$ (for
spherical accretion of a hydrogen-rich gas), the {X}-ray luminosity
exceeds the so-called Eddington limit at which the radiation pressure force on the accreting matter
exceeds the gravitational attraction force of the compact star (Davidson \& Ostriker~1973).
The concept of an Eddington accretion limit also applies to black holes.
At accretion rates $>\dot{M}_{\rm Edd}$ the excess accreting matter will pile up around the compact object
and form a cloud optically thick to {X}-rays, thus quenching the source. Therefore, in the observed
`persistent' LMXBs and HMXBs the accretion rates must be in the range $10^{-11}-10^{-8}\,M_{\odot}$~yr$^{-1}$.
However, the {\it mass transfer rate} from the companion star towards the neutron star may be considerably
larger. Calculations show that mass transfer exceeding $10^{-4}\,M_{\odot}$~yr$^{-1}$ may still be
dynamically stable (see Fig.~\ref{IMXB}). In cases where super-Eddington mass transfer occur, the
excess material is ejected in a jet -- e.g. as observed in the source SS433 (see the chapter by Fender in this book
and King \& Begelman 1999). 

\subsection{Accretion-induced magnetic field decay}
In the past decade it has been possible --  under certain assumptions -- to compute the accretion-induced 
magnetic field decay of a binary neutron star -- see e.g.
Geppert \& Urpin~(1994), Konar \& Bhattacharya~(1997) and Cumming, Zweibel \& Bildsten~(2001).
It is often assumed that the magnetic field has been generated in the outer
crust by some unspecified mechanism, e.g. by thermomagnetic effects
(Blandford, Applegate \& Hernquist 1983), during or shortly after
the neutron star is formed.
The electrical conductivity in the neutron star crust is mainly a function of
mass-density, temperature and degree of lattice impurities. By combining the conductive properties with a
cooling model, as well as a temperature profile, one can
calculate the evolution of the magnetic field by solving the
induction equation:
\begin{equation}
\frac{\partial \vec B}{\partial t} = - \frac{c^2}{4 \pi} \vec \nabla \times
(\frac{1}{\sigma_{\rm el}} \times \vec \nabla \times \vec B)
 + \vec \nabla \times (\vec v \times \vec B)     \label{emhd}
\end{equation}
where $\vec v$ is the velocity of accreted material movement and $\sigma_{\rm el}$ is the electrical conductivity of
the medium. By choosing a simple vector potential, and introducing the so-called Stokes' stream function,
the above equation reduces to a second order partial differential equation, which can be solved numerically.
The current distribution, which is responsible for the $B$-field,
migrates inward as a result of diffusion and enters the highly conducting parts
of the neutron star. In that inner region the electrical conductivity is very high (superconducting) and hence
the magnetic field will essentially be stable forever -- it freezes out at
a residual value.
The calculated effect of the ohmic dissipation results in final $B$-fields of $10^8 \sim 10^9$~G corresponding
to the values estimated from observed millisecond pulsars. 
A very different model, in which it is assumed that the field is anchored in the superconducting and superfluid
neutron star core, was put forward by Ruderman~(1998) and collaborators. In this physically very elegant model,
the field is driven out of the core by rotational slow-down, and also final fields of $10^8 \sim 10^9$~G result. 
See also the reviews by Bhattacharya \& Srinivasan~(1995) and Bhattacharya~(2002)
for further details and alternative models. 

In the case of crustal field decay models
the next important step would be to solve the evolution self-consistently. That is, to use good stellar
evolution models to calculate the mass transfer rate of the donor star, determine the orbital dynamical response,
check whether the material is accreted onto the neutron star or not, determine the crustal temperature
from the nuclear burning of accreted material, re-calculate the electrical conductivity and its effect
on the $\vec{B}$-field. The latter then affects the Alfv\'en radius, and thus the accretion rate and spin period,
which again influences on the orbital evolution and response of the donor star, etc.
However, our state of understanding the magnetic field decay mechanisms, and of mass and angular momentum loss
mechanisms during binary evolution, is still too fragmentary too allow for a detailed quantitative calculation
of the evolution of a rotating magnetized neutron star in a binary system.

\section*{Acknowledgments}
  T.M.T gratefully acknowledges support from the Danish Natural Science Research Council
  under grant no.$\,$56916.


\end{document}